\documentclass[preprint2]{aastex63}

\usepackage{CJK}

\newcommand{\petit}{\texttt{petitRADTRANS}}

\newcommand{\pmn}{\texttt{PyMultiNest}}

\newcommand{\teff}{T$_{\rm eff}$}

\newcommand{\caltech}{Department of Astronomy, California Institute of Technology, Pasadena, CA 91125, USA}
\newcommand{\gps}{Division of Geological \& Planetary Sciences, California Institute of Technology, Pasadena, CA 91125, USA}
\newcommand{\ucsc}{Department of Astronomy \& Astrophysics, University of California, Santa Cruz, CA95064, USA}
\newcommand{\keck}{W. M. Keck Observatory, 65-1120 Mamalahoa Hwy, Kamuela, HI, USA}
\newcommand{\ucla}{Department of Physics \& Astronomy, 430 Portola Plaza, University of California, Los Angeles, CA 90095, USA}
\newcommand{\jpl}{Jet Propulsion Laboratory, California Institute of Technology, 4800 Oak Grove Dr.,Pasadena, CA 91109, USA}
\newcommand{\ucsd}{Center for Astrophysics and Space Sciences, University of California, San Diego, La Jolla, CA 92093}
\newcommand{\ifahonolulu}{Institute for Astronomy, University of Hawai`i, 2680 Woodlawn Drive, Honolulu, HI 96822, USA}

\received{\today}
\revised{}
\accepted{}
\submitjournal{ApJ}

\shorttitle{C/O HR 7672 AB}
\shortauthors{Wang et al.}
\graphicspath{{./}{figures/}}

\begin{document}
\begin{CJK*}{UTF8}{gbsn}

\title{Retrieving the C and O Abundances of HR 7672~AB: a Solar-Type Primary Star with a Benchmark Brown Dwarf}

\correspondingauthor{Ji Wang}
\email{wang.12220@osu.edu}

\author[0000-0002-4361-8885]{Ji Wang (王吉)}
\affiliation{Department of Astronomy, The Ohio State University, 100 W 18th Ave, Columbus, OH 43210 USA}

\author{Jared R. Kolecki}
\affiliation{Department of Astronomy, The Ohio State University, 100 W 18th Ave, Columbus, OH 43210 USA}

\author[0000-0003-2233-4821]{Jean-Baptiste Ruffio}
\affiliation{\caltech}

\author[0000-0003-0774-6502]{Jason J. Wang (王劲飞)}
\altaffiliation{51 Pegasi b Fellow}
\affiliation{\caltech}

\author{Dimitri Mawet}
\affiliation{\caltech}
\affiliation{\jpl}

\author{Ashley Baker}
\affiliation{\caltech}

\author{Randall Bartos}
\affiliation{\jpl}

\author{Geoffrey A. Blake}
\affiliation{\gps}

\author{Charlotte Z. Bond}
\affiliation{UK Astronomy Technology Centre, Royal Observatory, Edinburgh EH9 3HJ, United Kingdom}

\author{Benjamin Calvin}
\affiliation{\caltech}
\affiliation{\ucla}

\author{Sylvain Cetre}
\affiliation{\keck}

\author[0000-0001-8953-1008]{Jacques-Robert Delorme}
\affiliation{\keck}
\affiliation{\caltech}

\author{Greg Doppmann}
\affiliation{\keck}

\author{Daniel Echeverri}
\affiliation{\caltech}

\author[0000-0002-1392-0768]{Luke Finnerty}
\affiliation{\ucla}

\author[0000-0002-0176-8973]{Michael P. Fitzgerald}
\affiliation{\ucla}

\author[0000-0001-5213-6207]{Nemanja Jovanovic}
\affiliation{\caltech}

\author{Michael C. Liu}
\affiliation{\ifahonolulu}

\author{Ronald Lopez}
\affiliation{\ucla}



\author{Evan Morris}
\affiliation{\ucsc}

\author[0000-0002-8823-8237]{Anusha Pai Asnodkar}
\affiliation{Department of Astronomy, The Ohio State University, 100 W 18th Ave, Columbus, OH 43210 USA}

\author{Jacklyn Pezzato}
\affiliation{\caltech}

\author{Sam Ragland}
\affiliation{\keck}

\author[0000-0001-8127-5775]{Arpita Roy}
\affiliation{Space Telescope Science Institute, 3700 San Martin Drive, Baltimore, MD 21218, USA}
\affiliation{Department of Physics and Astronomy, Johns Hopkins University, 3400 N Charles St, Baltimore, MD 21218, USA}

\author[0000-0003-4769-1665]{Garreth Ruane}
\affiliation{\caltech}
\affiliation{\jpl}

\author{Ben Sappey}
\affiliation{\ucsd}

\author{Tobias Schofield}
\affiliation{\caltech}

\author{Andrew Skemer}
\affiliation{\ucsc}

\author{Taylor Venenciano}
\affiliation{Physics and Astronomy Department, Pomona College, 333 N. College Way, Claremont, CA 91711, USA}

\author[0000-0001-5299-6899]{J. Kent Wallace}
\affiliation{\jpl}

\author{Nicole L. Wallack}
\affiliation{\gps}

\author{Peter Wizinowich}
\affiliation{\keck}

\author{Jerry W. Xuan}
\affiliation{\caltech}



\begin{abstract}


A benchmark brown dwarf (BD) is a BD whose properties (e.g., mass and chemical composition) are precisely and independently measured. Benchmark BDs are valuable in testing theoretical evolutionary tracks, spectral synthesis, and atmospheric retrievals for sub-stellar objects. Here, we report results of atmospheric retrieval on a synthetic spectrum and a benchmark BD---HR 7672~B---with \petit. First, we test the retrieval framework on a synthetic PHOENIX BT-Settl spectrum with a solar composition. We show that the retrieved C and O abundances are consistent with solar values, but the retrieved C/O is overestimated by 0.13-0.18, which is $\sim$4 times higher than the formal error bar. Second, we perform retrieval on HR 7672~B using high spectral resolution data (R=35,000) from the Keck Planet Imager and Characterizer (KPIC) and near infrared photometry. We retrieve [C/H], [O/H], and C/O to be $-0.24\pm0.05$, $-0.19\pm0.04$, and $0.52\pm0.02$. These values are consistent with those of HR 7672~A within 1.5-$\sigma$. As such, HR 7672~B is among only a few benchmark BDs (along with Gl 570~D and HD 3651~B) that have been demonstrated to have consistent elemental abundances with their primary stars. Our work provides a practical procedure of testing and performing atmospheric retrieval, and sheds light on potential systematics of future retrievals  using high- and low-resolution data.  

\end{abstract}



\section{Introduction}
\label{sec:intro}

Physical and chemical properties of sub-stellar objects can be inferred by modeling observed spectra. When the spectral modeling contains the key physics and chemistry in a self-consistent way, we call it a forward-modeling approach. When the spectral modeling uses a flexible parameterization without a rigorous and self-consistent treatment of involoved physics and chemistry, we call it a free-retrieval approach. Both approaches are valuable in the study of atmospheres of exoplanets. 

The majority of forward modeling and free retrieval analyses have been applied to transiting planets~\citep[e.g., ][]{Waldmann2015,Zhang2020}, see also a recent review by~\citet{Madhusudhan2019}. Fewer such analyses exist for directly-imaged exoplanets~\citep[e.g., ][]{Konopacky2013}, mainly because of the smaller number of targets. However, direct imaging of exoplanets is considered a top science goal for future extremely large telescopes~\citep[ELTs, e.g., ][]{Mawet2019} and NASA missions such as HabEx and LUVOIR~\citep{Gaudi2021}. Toward the goals of: (1) better understanding exoplanets' atmospheres, and (2) ultimately detecting biosignatures via spectroscopic observations, we need to address the following challenges facing direct imaging of exoplanets. 

First, while most spectral modeling frameworks~\citep[e.g., ][]{Molliere2019,Lavie2017,Baudino2017} are bench-marked against each other, they have not been tested against benchmark brown dwarfs (BDs), for which we know the dynamical mass (e.g., from radial velocity and/or astrometric measurements) and the chemical composition (e.g., from their companion primary stars). The chemical homogeneity assumption---that stars and BDs form within the same molecular cloud should have identical chemical compositions---has been tested and applied for solar-type and M-type companions~\citep[e.g., ][]{Mann2013,Nelson2021}. Such benchmarking efforts have only been done using low spectral resolution data for very limited number of BDs~\citep[GJ 570 D and HD 3651 B, ][]{Line2015}.

Second, future data from ELTs are likely to include those with very high spectral resolution (R$\sim$100,000), e.g., MODHIS at TMT~\citep{Mawet2019} and METIS at E-ELT~\citep{Brandl2021}, but few existing spectral modeling frameworks can handle such high-resolution data. This drawback has been realized and recent advances have been made for transiting planets~\citep[e.g., ][]{Fisher2020}.   

Third, combining high- and low-resolution data presents a challenge. While theoretical frameworks have been proposed~\citep[e.g., ][]{Brogi2017} and applied to transiting planets~\citep{Gandhi2019}, no such spectral modeling framework has been tested using benchmark BDs to better characterize directly-imaged exoplanets. 

The above challenges motivate the retrieval framework in this paper, which can be viewed as an extension of \petit~\citep{Molliere2019}. In this paper, we (1) test the framework against a synthetic PHOENIX spectrum with known solar composition; and (2) apply the framework on a benchmark BD---HR 7672 B---whose stellar abundance is inferred from its G-type primary star HR 7672 A. We first provide an overview of recent progress in the spectral modeling of BDs and the nature of the HR 7672 AB system. 

\subsection{Recent Progress in Modeling BDs}

~\citet{Line2015} presented ground-breaking work where they showed that two benchmark T-type BDs (Gl 570 D and HD 3651 B) have similar C, O, and Fe abundances as their primary stars. T-type BDs were chosen because their cloudless atmospheres, which are simpler than cloudy atmospheres and require fewer modeling parameters. However, ~\citet{Maire2020} found evidence of patchy or thin cloud for a T-type benchmark BD, HD 19467 B. By comparing to a theoretical model grid and a BD spectral library, ~\citet{Rickman2020} studied another benchmark BD (HD 13724 B) but provided no C or O abundance. ~\citet{Zhang2020} used a forward modeling approach to study three late-T-type benchmark dwarfs, HD 3651 B, GJ 570 D, and Ross 458 C and found discrepencies in tempeature, radius, and surface gravity, which they attributed to clouds, reduced vertical temperature gradients, or disequilibrium processes.~\citet{Kitzmann2020} used \texttt{Helios-r2} to retrieve the atmospheric properties of GJ 570 D and stressed the impact of chemical equilibrium on inferred C and O abundances.

There are more challenges in the L-type BD regime because of clouds.~\citet{Burningham2017} performed retrieval analyses on two BDs: 2MASS J05002100+0330501 and 2MASSW J2224438−015852 and found evidence of cloudy conditions, but failed to address the anomalously high CO abundance, which was attributed to unrecognized shortcomings in their retrieval model. ~\citet{Gonzales2020} conducted a similar study on the L- and T-type BD binary J14162408+1348263AB and found a consistent C/O ratio between the binary pair. ~\citet{Peretti2019} studied another L-type benchmark BD, HD 4747B. However, their retrieved C and O abundances were off by 1.0 dex and 0.4 dex ($x$ dex corresponds to a factor of 10$^x$) from the primary star, which is a $\sim$2-$\sigma$ discrepancy.   

\subsection{HR 7672 A and B}

HR 7672 A is a solar-type G0 star~\citep{Brewer+2016}. Due to the radial velocity trend of HR 7672 A, ~\citet{Liu2002} detected HR 7672 B, which is a L-type BD companion at a separation of 0.8$^{\prime\prime}$. The estimated effective temperature was $1510-1850$ K. Apparent $J$, $H$, and $K$-band magnitudes were reported in~\cite{Boccaletti2003} with values of $\sim$14.4, 14.04$\pm$0.14, and 13.04$\pm$0.10.~\citet{Crepp2012} refined the dynamical mass of HR 7672 B to lie between 65.6 and 71.1 M$_{\rm{Jupiter}}$ using combined radial velocity and astrometric data sets. With improved Gaia astrometry and a longer RV baseline,~\citet{Brandt2019} further constrained the mass of HR 7672 B to be 72.7$\pm$0.8 M$_{\rm{Jupiter}}$. The rich literature on HR 7672 AB and the recently obtained high-resolution spectrum for HR 7672 B make the system an ideal benchmark to test retrieval frameworks. 

\begin{figure*}[ht]
   \centering
\begin{tabular}{c}
\includegraphics[width=0.7\paperwidth]{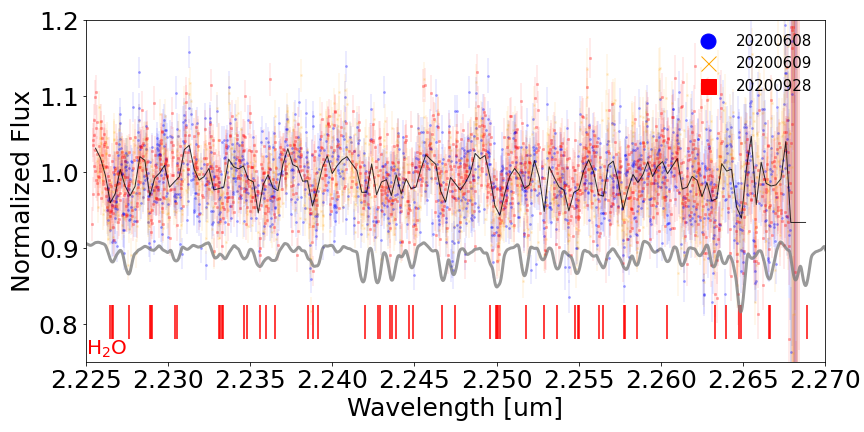}\\
\includegraphics[width=0.7\paperwidth]{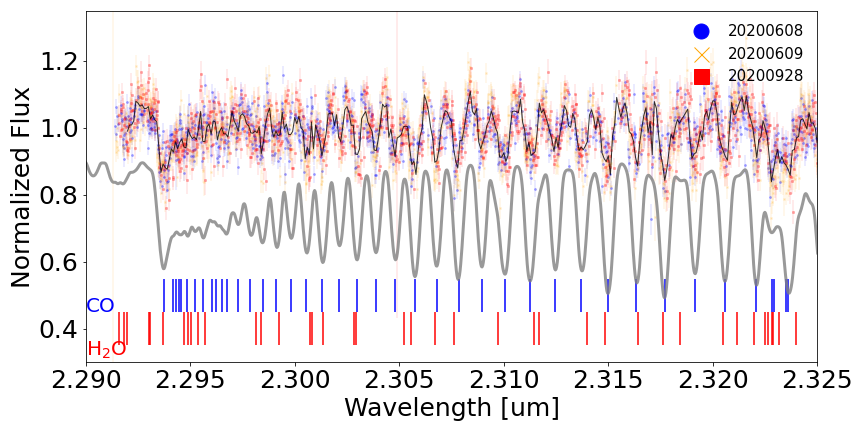}\\
\end{tabular}
\caption{KPIC data were obtained on UT 06/08/2020 (blue circles), 06/09/2020 (yellow crosses), and 09/28/2020 (red squares) and we use two spectral orders (top and bottom panels) for the analyses in this paper. Black lines through the data points are binned spectra with a bin size of 1 $\AA$. For comparison, a PHOENIX BT-Settl spectrum with a T$_{\rm{eff}}$ of 1600 K, $\log$(g) of 5.5, and v$\sin i$ of 40 km$\cdot s^{-1}$ is shown in grey. Prominent CO and H$_2$O line locations are marked as blue and red vertical lines. 
\label{fig:dataset}}
\end{figure*} 

The paper is organized as follows. \S \ref{sec:observation} summarizes the observational data. We conduct an independent abundance analysis for HR 7672 A and in report the results in \S \ref{sec:st_ab}. \S \ref{sec:data_reduction} describes our framework of ultra-cool atmosphere retrieval. In \S \ref{sec:test}, we test the framework using a PHOENIX BT-Settl spectrum with solar composition. In \S \ref{sec:combined_retrieval}, we apply the framework to HR 7672 B to check if the retrieved abundances are consistent with the measured abundances from HR 7672 A. Finally, we summarize the paper in \S \ref{sec:summary}.

\section{Data}
\label{sec:observation}

We used archival high-resolution spectra to determine the primary stellar abundances. For the BD atmospheric abundances, we obtained high-resolution (R=35,000) data using NIRSPEC fed by the Keck Planet Imager and Characterizer~\citep[KPIC, ][]{Mawet2018, Jovanovic2019, Delorme2020}. To supplement the spectroscopic data, we also gathered photometric data for HR 7672 B. The photometric data serve as low-resolution data (R$\sim$3-5) in our joint high- (R$\sim$35,000) and low-resolution retrieval, and our retrieval code can handle any spectral resolution R between a few and 1,000,000.      

\subsection{Spectroscopic Data for HR 7672 A}
\label{sec:spec_data_a}

Keck/HIRES spectroscopic data with wavelength coverage ranging from 4350\AA-8690\AA~was retrieved from the Keck Observatory Archive (KOA). These observations were taken on July 5, 2018 between 13:24:13-13:25:24 UTC (PI: Yong).

Using the associated calibration files, this raw data was then reduced into a 1-dimensional spectrum using the MAKEE pipeline\footnote{\url{https://sites.astro.caltech.edu/~tb/makee/}}. The MAKEE reduction process involves wavelength scaling using calibration images (ThAr arcs), flat-field  correction for large-scale continuum fluctuations, and CCD bias subtraction. The resulting signal-to-noise ratio at 7770\AA~(near the \ion{O}{1} triplet) is approximately 300.

\subsection{Spectroscopic Data for HR 7672 B}
\label{sec:spec_data_b}


HR 7672 B was observed in K-band ($R\sim35,000$) with Keck/NIRSPEC/KPIC on three occasions summarized in Table \ref{tab:KPICHR7672B}. An A0 standard star, zet Aql, was observed each epoch to calibrate the combined transmission of the atmosphere and the instrument. The data were acquired and reduced following the approach described in ~\citet{Delorme2021}. Images were first background subtracted and bad pixel corrected. The fiber trace locations and widths were calibrated using the standard star. The spectra were then extracted using optimal extraction~\citep{Horne1986}. The wavelength solution was derived from observations of the M-giant HIP 81497 using a forward model of the tellurics generated with the Planetary Spectrum Generator  \citep{Villanueva2018} and a Phoenix model of the star ($log(g/[1\,\mathrm{cm.s}^{-2}])=1$; $T_{\mathrm{eff}}=3600\,\mathrm{K}$ \citet{Husser2013}).

To obtain the normalized spectra, we perform the following procedures. First, we subtract background from both the target star and the telluric standard star. Second, we divide the target star spectrum by the telluric standard star spectrum to remove the blaze function and telluric lines. Third, we normalize the target star spectrum by dividing by the median of the spectrum. The median spectrum is obtained by running a median kernel with width of 200 pixels. The kernel width, which is about one tenth of the entire spectral order, is selected in order not to affect the molecular absorption lines/bands. Finally, we shift the normalized spectra by a certain radial velocity. The radial velocity is determined by cross-correlating the spectrum of each date with a PHOENIX synthetic spectrum~\citep[][and references therein]{Allard2012, Allard2013, Baraffe2015} with T$_{\rm{eff}}$ of 1600 K, $\log$(g) of 5.5, and solar abundances.  

KPIC data sets for HR 7672 B are shown in Fig. \ref{fig:dataset}. We use 2 (out of 9) spectral orders centering around 2.25 and 2.31 $\mu$m. For other orders, we cannot calibrate the wavelength, the signal-to-noise-ratio is too low, these orders are heavily contaminated by telluric CO$_2$ lines.

\begin{table}
  \centering
\begin{tabular}{ |c|ccc| } 
 \hline
  Object & Date & Exposure time & Note\\ 
 \hline\hline
zet Aql & 2020-06-08 & $3\times10$ sec & standard \\
HR 7672 B & 2020-06-08 & $11\times10$ min & \\
zet Aql & 2020-06-09 & $3\times10$ sec & standard \\
HR 7672 B & 2020-06-09 & $10\times10$ min & \\
zet Aql & 2020-09-28 & $4\times4.4$ sec & standard \\
HR 7672 B & 2020-09-28 & $7\times10$ min & \\
 \hline
\end{tabular}
\caption{K-band observations of HR 7672 A and B with KPIC.}
\label{tab:KPICHR7672B}
\end{table}

\subsection{Photometric Data for HR 7672 B}
\label{sec:phot_data_b}

We used $J$, $H$, and $K$-band differential magnitudes~\citep{Crepp2012} to convert to apparent magnitudes. In the conversion, magnitudes from HR 7672 A are from SIMBAD. The apparent magnitudes are then converted to physical flux in the unit of W$\cdot\mu$m$^{-1}\cdot$m$^{-2}$, assuming a distance of $17.72\pm0.02$ pc~\citep[Gaia DR2, ][]{Prusti2016, Brown2018}. As such, we adopt $1.5\times10^{-15}$, $1.8\times10^{-15}$, and $1.4\times10^{-15}$ W$\cdot\mu$m$^{-1}\cdot$m$^{-2}$ for $J$, $H$, and $K$-band, respectively. The fractional uncertainties for $J$, $H$, and $K$-band fluxes are measured to be 20\%, 12\%, and 9\%~\citep{Crepp2012}. {Note that the $J$-band flux and the associated error are likely to be affected by speckle contamination~\citep{Boccaletti2003}. } 

\begin{figure}[h]
\includegraphics[width=8.0cm]{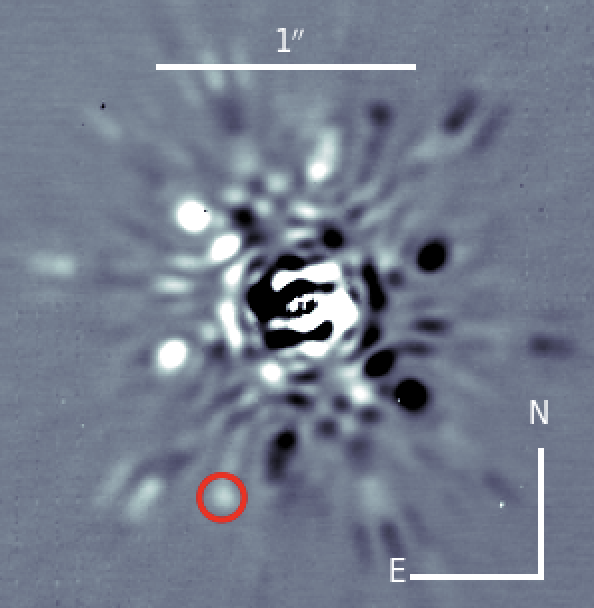}
\caption{$L'$-band image of HR 7672 B (indicated by the red circle). The image is shown in linear scale with saturation at 10\% of the maximum value. 
\label{fig:l_band_img}}
\end{figure} 

We measured the $L'$-band of HR 7672 B using Keck/NIRC2 imaging data from the KOA taken on October 27, 2002 (PI: Graham). We used a total of 40 frames with 0.018~s exposures and 500 coadds. The images were sky-subtracted using sky images taken after the sequence. Because the images were taken in position angle mode, no differential imaging techniques could be used. Instead, we registered all of the frames together by fitting the star in each image with a 2D Gaussian and combined all of the images together. HR 7672 B fortunately lies between the diffraction spikes from the primary star, so is visible after simple mean combination of frames (Fig. \ref{fig:l_band_img}). To moderately improve the detection, we subtracted a copy of the image that was rotated by 180 degrees to remove symmetric features in the stellar PSF. {We then fit a 2D Gaussian to the brown dwarf companion, and measured a flux ratio of $(1.3 \pm 0.3) \times 10^{-3}$ in $L'$-band. This leads to a $L$-band flux of $(5.5 \pm 1.3) \times10^{-16}$ W$\cdot\mu$m$^{-1}\cdot$m$^{-2}$. }

\section{Stellar Atmospheric Parameters and Abundances of HR 7672 A}
\label{sec:st_ab}

We performed our stellar parameter derivation, abundance measurements, and error analysis using the same code and 1D-LTE analysis process explained in Sections 5.1 through 5.3 of \citet{Kolecki+2021}. We present a brief overview here. The results are reported in Table \ref{table:Params} and Table \ref{tab:CO_abundances}. 

\subsection{Stellar Parameters}
\subsubsection{Efffective Temperature and Surface Gravity}
We derived \teff\ and $\log$(g) by comparing magnitudes from Gaia ($G$, $Bp$, $Rp$), 2MASS ($J$, $H$, $Ks$), and WISE ($W1$, $W2$, $W3$, $W4$) with theoretical stellar isochrones from the Dartmouth Stellar Evolution Database\footnote{\url{http://stellar.dartmouth.edu/models/}} \citep{Dotter+2007}. 

At a fixed metallicity, we calculated the sum of the differences (i.e. residuals) between observed and isochronal values for each pass band magnitude along an age-EEP (equivalent evolutionary phase) isochrone grid. A contiguous area where the residuals are within 10\% of their absolute minimum on the grid is chosen as valid points on this isochrone grid to extract \teff\ and $\log$(g) values from. We took the mean of all these resulting values as the final parameters.

\subsubsection{Microturbulence ($\xi$)}
To determine a value for the microturbulence velocity ($\xi$), we first fixed \teff\ and $\log$(g), then minimized the slope of the correlation between abundance and reduced equivalent width (REW) by calculating the slope for a grid of microturbulence velocities and interpolating to find the velocity where the slope was equal to 0. 

Uncertainty in the microturbulence parameter was determined by perturbing the value until the slope fell outside the range of uncertainty defined by the linear regression fit.

\subsection{Abundances}

The use of \texttt{abfind} in MOOG~\citep{Sneden1973} requires measurement of the equivalent widths of spectral absorption lines. This was done using a semi-automated program. This program displays measurements from fits of both a Gaussian profile and a Voigt profile to each line, along with its direct measurement of the observed data. 

From these three options we chose the method which most closely fit trends in the data. This manual screening process allows for the most accurate measurements to be kept for all lines, mitigating the effects of noise, contamination, and improper line fitting. If these effects were too great, we discarded the line from the final line list.

To mitigate the effects of strong line damping on our results, we chose a maximum equivalent width of 100 m\AA, and ignored all lines stronger than this cutoff.

Uncertainties in all abundances were calculated according to the method outlined in \citet{Epstein+2010}. This method takes the base uncertainty to be the standard deviation of the mean abundance, and modifies it to account for uncertainty and covariances in the atmospheric parameters.

Wavelength, excitation potential and log(gf) data for Fe, C, and O lines was supplied by the NIST Atomic Spectra Database \citep{NIST_ASD}. Solar abundances were taken from \citet{Palme2014}.

\subsubsection{Iron Abundances and [Fe/H]}

Once the iteration of stellar parameters was completed, we determined initial [\ion{Fe}{1}/H] and [\ion{Fe}{2}/H] values of $-0.07 \pm 0.07$ dex and $-0.06 \pm 0.08$ dex respectively, relative to the adopted solar iron abundance of 7.48~\citep{Palme2014}. 

From these values, we aimed to use correction tables to account for the limitations of a 1D-LTE (1-dimensional local thermodynamic equilibrium) analysis. \citet{Amarsi+2019} have calculated a grid of 3D-LTE/1D-LTE corrections for several \ion{Fe}{2} lines which we have used to adjust our \ion{Fe}{2} abundance measurement. 

Using the point on the grid (\teff = 6000 K, $\log$(g) = 4.5, $\xi$ = 1.0 km/s, [Fe/H] = 0.00) that most closely matches the 1D-LTE parameters of HR 7672, we found that the average correction for the \ion{Fe}{2} lines we measured results in an increase of this abundance by 0.05 dex. Thus, we report our final [\ion{Fe}{2}/H] = $-0.01 \pm 0.08$.

Taking the overall metallicity ([Fe/H]) to be equal to the average of the \ion{Fe}{1} and \ion{Fe}{2} abundances and accounting for the effect this has on the uncertainty, [Fe/H] = $-0.04 \pm 0.07$.

\subsubsection{LTE C and O Abundances}

We then proceeded to derive the abundances of carbon and oxygen. The oxygen abundance was derived from the \ion{O}{1} 777 nm triplet feature, while carbon abundance was derived from lines at 5380.32, 6587.62, 7113.17, 7115.17, and 7116.96\AA. From these line features, we found LTE abundances of log($\epsilon_C$) = $8.42 \pm 0.05$ and log($\epsilon_O$) = $8.88 \pm 0.06$, where $\epsilon_X$ is the abundance for element $X$ and $\epsilon_H$ is defined at 10$^{12}$. The C and O abundances correspond to [C/H] and [O/H] of $-0.08 \pm 0.05$~dex and $0.15 \pm 0.06$~dex respectively.

\subsubsection{Consideration of Non-LTE Effects on C and O Abundances}
As the abundance derived from the \ion{O}{1} triplet is known to be affected significantly by the LTE assumption, we used the grid of 3D non-LTE (NLTE) corrections calculated by \citet{Amarsi+2019} to account for this.

Taking the point on this grid (\teff = 5999 K, $\log$(g) = 4.5, $\xi$ = 1.0 km/s, [Fe/H] = 0.00, log($\epsilon_C$) = 8.43, log($\epsilon_O$) = 8.89) that most closely matches the LTE parameters of HR 7672, we find that the NLTE corrections lower [O/H] by 0.22 dex and [C/H] by 0.01 dex. We note that while the carbon NLTE correction is within the uncertainty of [C/H], it does affect the final C/O ratio.

After the NLTE corrections were applied, our final C and O abundances were determined to be log($\epsilon_C$) = $8.41 \pm 0.05$ and log($\epsilon_O$) = $8.66 \pm 0.06$, corresponding to [C/H] and [O/H] of $-0.09 \pm 0.05$~dex and $-0.07 \pm 0.06$~dex.

This leads to an adopted C/O = $0.56\pm0.11$, where the uncertainty of C/O is given by adding in quadrature the fractional uncertainty of the numerical quantities of C and O and multiplying the result by the C/O value.

\subsection{Comparing to Previous Work}
HR 7672 A was included in the abundance analysis samples of \citet{daSilva2015}, \citet{Brewer+2016}, and \citet{Luck2017}, though the first did not include a [O/H] measurement. Each derived an effective temperature within 30 K of the value adopted in this paper, within the 1$\sigma$ uncertainty range. Each also derived a $\log$(g) value within 0.03 dex of our photometric measurement.

Our [Fe/H] measurement is consistently below but in agreement with those of the literature by $\sim$1$\sigma$. A similar consistency level is found in the carbon and oxygen abundances. Our log($\epsilon_C$) is within 1$\sigma$ of the values derived by \citet{Brewer+2016} and \citet{Luck2017}, and within 2$\sigma$ of the value drived by \citet{daSilva2015}. Our log($\epsilon_O$) is consistent to within 1$\sigma$ with both mentioned papers which also measured the abundance of oxygen. For more information, values from each paper are compared with our results in Tables \ref{table:Params} and \ref{tab:CO_abundances}.

We note that \citet{Luck2017} does not provide values for uncertainties of C and O, but does provide error bars in plots of these chemical abundances (Figure 13 of that paper). We compared the length of these bars with the scale of the axes in order to get numerical values for use in Tables \ref{table:Params} and \ref{tab:CO_abundances}.

\section{Atmosphere Modeling and Retrieval for HR 7672 B}
\label{sec:data_reduction}

\subsection{Overview of the Modeling and Retrieval Framework}

Our framework to model exoplanet atmospheres based on \petit\ is described in~\citet{Wang2020} (WW20 hereafter). Since low-resolution broad-band data and high-resolution spectroscopic data are used in the retrievals, we consider both resolution modes in \petit~(R=1,000 and R=1,000,000). To sample the posterior distribution in a Bayesian framework, we used \pmn~\citep{Buchner2014} which is based on the MultiNest sampling algorithm~\citep{Feroz2009}. 

Since \petit\ is the core of our retrieval work, the framework described in this paper can be viewed as an extension of \petit. Here, we demonstrate that our framework can analyze data sets that include both high- and low-spectral-resolution data and that our framework is benchmarked against synthetic and observed spectra. 


\subsection{Major Updates Since WW20}

We list below the updates of our retrieval code since WW20, which make the framework more versatile, flexible, and physical. 

\subsubsection{Considering High-Resolution Spectroscopic Data}

In WW20, we excluded the high-res mode in \petit~for practical reasons. First, computational time was greatly reduced. Second, the highest spectral resolution in our data set was at R=5,000 and can be down-sampled to R=1,000, which is the low-res mode in \petit, without a significant loss of information content. However, our KPIC data has a spectral resolution of R=35,000 and we need to invoke the high-res mode in \petit~in modeling BD spectra to maximize the information content. In practice, we compute R=1,000,000 modeled spectra in a very narrow spectral range from 2.18 to 2.36 $\mu$m, which covers the two spectral orders that we consider. We then downsample the data to match the spectral resolution of KPIC. 

\subsubsection{Flexible P-T Profile}
\label{sec:flexible_pt}

In WW20, we use an analytical P-T profile~\citep{Parmentier2014, Parmentier2015} to speed up the posterior sampling. However, we find evidence that the analytical P-T profile may not be sufficiently flexible and can bias the retrieved abundances. We therefore switch to a more flexible P-T profile~\citep{Piette2020}. In the newly-adopted P-T profile, there are eight variables including seven temperature differences at eight pre-defined pressure levels: 100.0, 33.3, 10.0, 3.3, 1.0, 0.1, 0.001, 0.00001 bar. The other variable is the temperature at 3.3 bar. 

\subsubsection{Physical Cloud Treatment}
\label{sec:cloud}

In WW20, we treated clouds as a grey opaque cloud with infinite opacity if pressures were higher than a certain threshold. In reality, however, this assumption of a grey opaque cloud may be inadequate. Using \petit, ~\citet{Molliere2020} adopted a more realistic treatment where silicate and iron clouds were considered. Here, we consider a silicate cloud that only consists of MgSiO$_3$ for the following reasons. First, it is found that an iron cloud is not as prominent as a silicate cloud~\citep{Gao2020} or the iron cloud is at a much deeper level than the MgSiO$_3$ cloud~\citep{Burningham2021}. Second, considering only MgSiO$_3$ rather than MgSiO$_3$ and Mg$_2$SiO$_4$ saves computational time, and using both silicate clouds is not expected to provide more meaningful and constraining results compared with a single-cloud model. 

In practice, we modify \petit~so that the cloud opacity is included for both low- and high-res modes. To make sure both modes handle cloud opacity consistently,
we consider two scenarios for the low-res mode: one with a MgSiO$_3$ cloud and the other one without a MgSiO$_3$ cloud (i.e., abundance for MgSiO$_3$ is set to be zero). We then interpolate the opacity difference of the two scenarios at the central wavelength of the high-res mode. The interpolated opacity is then added in the high-res mode in calculating modeled spectra. 

To find the cloud pressure, we intercept the MgSiO$_3$ condensation curve with the P-T profile. The intercepting pressure marks where the cloud deck is. Then the mass fraction of MgSiO$_3$ decays exponentially from the cloud deck as controlled by the f$_{\rm{sed}}$ parameter as discussed in \S \ref{sec:parametrization}.

\begin{figure*}[h!]
\begin{tabular}{l}
\includegraphics[width=17.0cm]{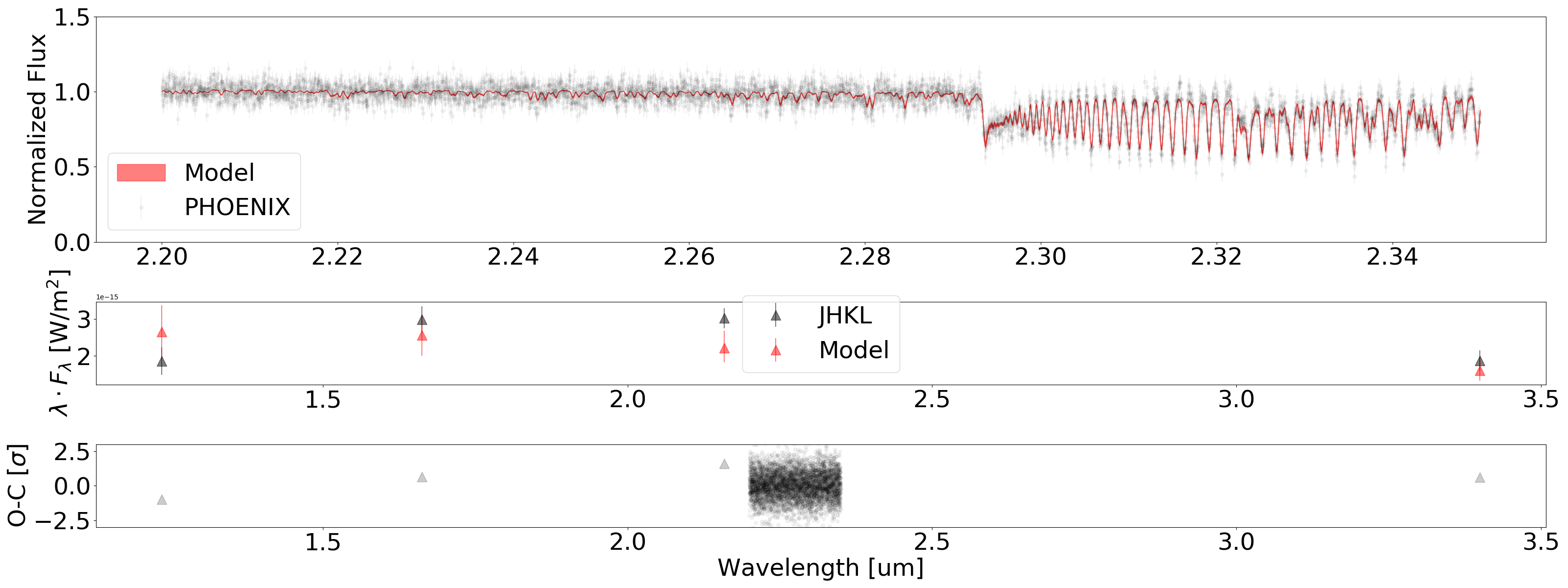} 
\end{tabular}
\caption{{\bf{Retrieved spectra based on the PHOENIX spectrum with a Gaussian prior (72.7$\pm$0.8 M$_{\rm{Jupiter}}$).}} Top two panels are simulated high-resolution spectroscopic and photometric data (black) and the modeled data (red). The bottom panel is a residual plot with data minus model and divided by errors. 
\label{fig:data_model_phoenix_fixed}}
\end{figure*} 

\begin{figure*}[h]
\begin{tabular}{l}
\includegraphics[width=17.0cm]{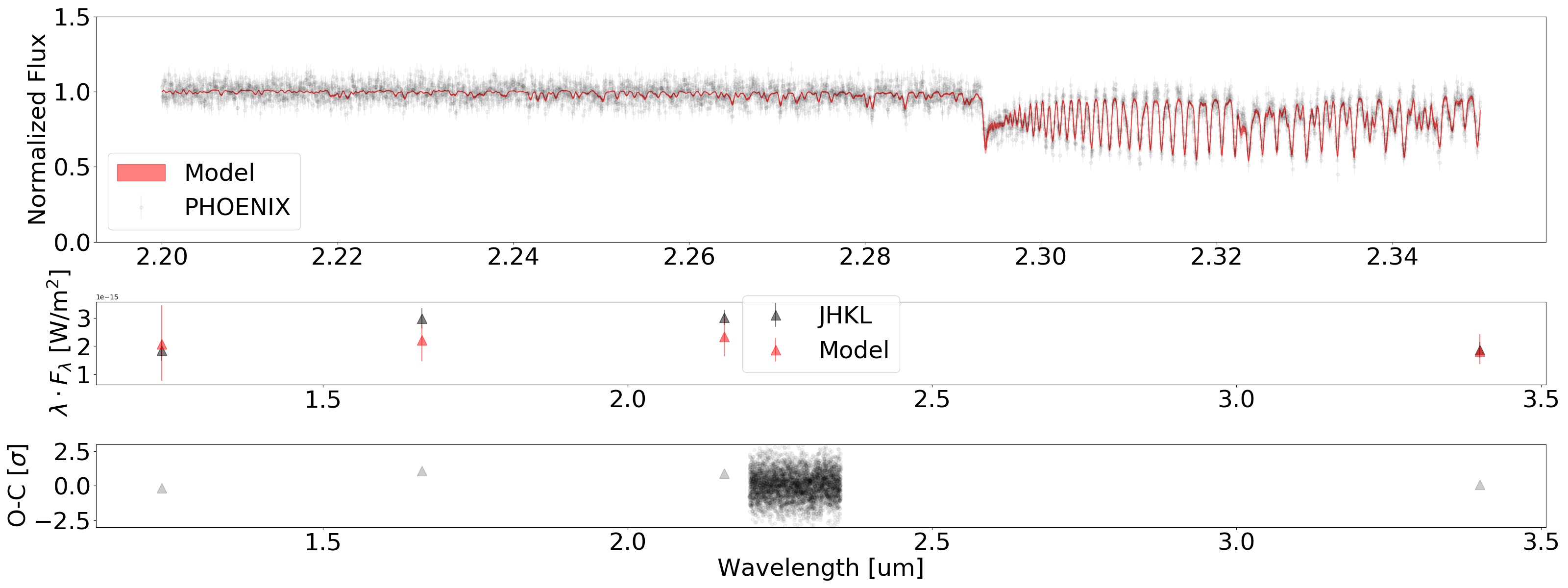}
\end{tabular}
\caption{{\bf{Retrieved spectra based on the PHOENIX spectrum with a flat mass prior between 10 and 100 M$_{\rm{Jupiter}}$.}} Top two panels are simulated high-resolution spectroscopic and photometric data (black) and the modeled data (red). The bottom panel is a residual plot with data minus model and divided by errors. 
\label{fig:data_model_phoenix_free}}
\end{figure*} 

\subsection{Parameterization}
\label{sec:parametrization}

There are 21 parameters in our retrieval code. Three parameters are used to described the BD properties: surface gravity ($\log$(g)), planet mass (m$_p$), and the projected rotational velocity (v$\sin i$). Four parameters are for the mass mixing ratio (MMR) of H$_2$O, CO, CO$_2$, and CH$_4$. {While there exist other species in the atmosphere, they do not significantly alter the spectroscopic and photometric observables. For example, PH$_3$ has rotation-vibrational features in $K$-band, but they are much weaker than those from CO and H$_2$O, even at a MMR that is a few orders of magnitude higher than expected at chemical equilibrium. Both TiO and VO are unaccounted in our model and may affect $J$-band photometry, but are estimated to be at the 0.1\% level, which is much smaller than the photometric uncertainty. Therefore, we focus on four molecular species: H$_2$O, CO, CO$_2$, and CH$_4$.} 

The MMRs are assumed to be constant at all pressures considered. The constant MMRs are justified by the narrow range of pressures that contribute to the thermal flux as well as the nearly constant MMRs for the two major C and O carriers H$_2$O and CO (see \S \ref{sec:chemical_eq}). Moreover, the constant MMR assumption is found to be favored over varying MMR as a function of altitude~\citep{Burningham2021}. Eight parameters are for the flexible P-T profile (\S \ref{sec:flexible_pt}). Similar to~\citet{Molliere2020}, we use four parameters to describe the cloud properties: MMR of MgSiO$_3$, vertical diffusion coefficient (K$_{zz}$), the ratio of the cloud particle settling and mixing velocities (f$_{\rm{sed}}$), and a log-normal particle size distribution parameter ($\sigma_g$). The other parameter is for a wavelength shift of the high-resolution spectrum between the data and model.


\subsection{Calculating Abundance Ratios}
In $K$-band, CO and H$_2$O lines are predominately present in the spectrum (Fig. \ref{fig:dataset}). Therefore, C and O abundances are mainly constrained by detecting and modeling CO and H$_2$O lines. For completeness sake, we also consider CO$_2$ and CH$_4$ in our spectral modeling.

We calculate C/H with the following equation:
\begin{equation}
\label{eq:extraction_c}
C/H = \frac{X_{\rm{CO}} + X_{\rm{CO_2}} + X_{\rm{CH_4}}}{2 \times X_{\rm{H_2}} + 4 \times X_{\rm{CH_4}} + 2 \times X_{\rm{H_2O}}},
\end{equation}
where X is volume mixing ratio (VMR). The conversion from MMR to VMR is given in WW20.

The MMRs for all considered species add up to unity. The molecular hydrogen to helium ratio is 3:1 from our primordial composition assumption. Similarly, O/H was calculated using the following equation:
\begin{equation}
\label{eq:extraction_o}
O/H = \frac{X_{\rm{CO}} + 2 \times X_{\rm{CO_2}} + X_{\rm{H_2O}}}{2 \times X_{\rm{H_2}} + 4 \times X_{\rm{CH_4}} + 2 \times X_{\rm{H_2O}}}.
\end{equation}
And C/O was calculated as:
\begin{equation}
\label{eq:extraction}
C/O = \frac{X_{\rm{CO}} + X_{\rm{CO_2}} + X_{\rm{CH_4}}}{X_{\rm{CO}} + 2 \times X_{\rm{CO_2}} + X_{\rm{H_2O}}}.
\end{equation}

\section{Testing with a PHOENIX BT-Settl Spectrum}
\label{sec:test}

Here we test our retrieval framework using a synthetic spectrum for which we know the C and O abundance. {The synthetic spectrum is from the PHOENIX BT-Settl model~\citep{Baraffe2015}. We choose a synthetic spectrum with T$_{\rm{eff}}$ of 1600 K, $\log$(g) of 5.5, and solar abundances\footnote{The fits file is available at: \url{https://phoenix.ens-lyon.fr/Grids/BT-Settl/CIFIST2011_2015/FITS/}}.} The effective temperature and surface gravity of the synthetic spectrum are similar to those of HR 7672 B.  

\subsection{Simulating the Data}

We obtain wavelength and flux from the PHOENIX spectrum and then scale the flux based on distance and radius. We use a distance of 17.72 pc and a radius of 0.75 R$_{\rm{Jupiter}}$. The radius is consistent with an object with $\log$(g) of 5.5 and 72 M$_{\rm{Jupiter}}$. The fluxes in $J$, $H$, $K$, and $L$-band are estimated and given the following fractional errors: 0.20, 0.12, 0.09, 0.15, which are the fractional errors for the actual photometric measurements for HR 7672 B. For fluxes, we use $1.5\times10^{-15}$, $1.8\times10^{-15}$, $1.4\times10^{-15}$, and $5.5\times10^{-16}$ W$\cdot\mu$m$^{-1}\cdot$m$^{-2}$ for $J$, $H$, $K$, and $L$-band, respectively. 

To simulate high-resolution spectroscopy data, we use a wavelength range from 2.20 to 2.35 $\mu$m. We apply a rotation broadening of 40 km$\cdot$s$^{-1}$, convolve the spectrum with a Gaussian kernel that corresponds to spectral resolution of R=35,000, and resample the spectrum with a sampling rate of $3\times10^{-5}$ $\mu$m, which translates to $\sim$2 pixels per resolution element. We add a randomized fractional error of 5\% to each data point. The 5\% fractional error is comparable to that in the actual HR 7672 B data. 

\begin{figure}[h]
\begin{tabular}{l}
\includegraphics[width=8.0cm]{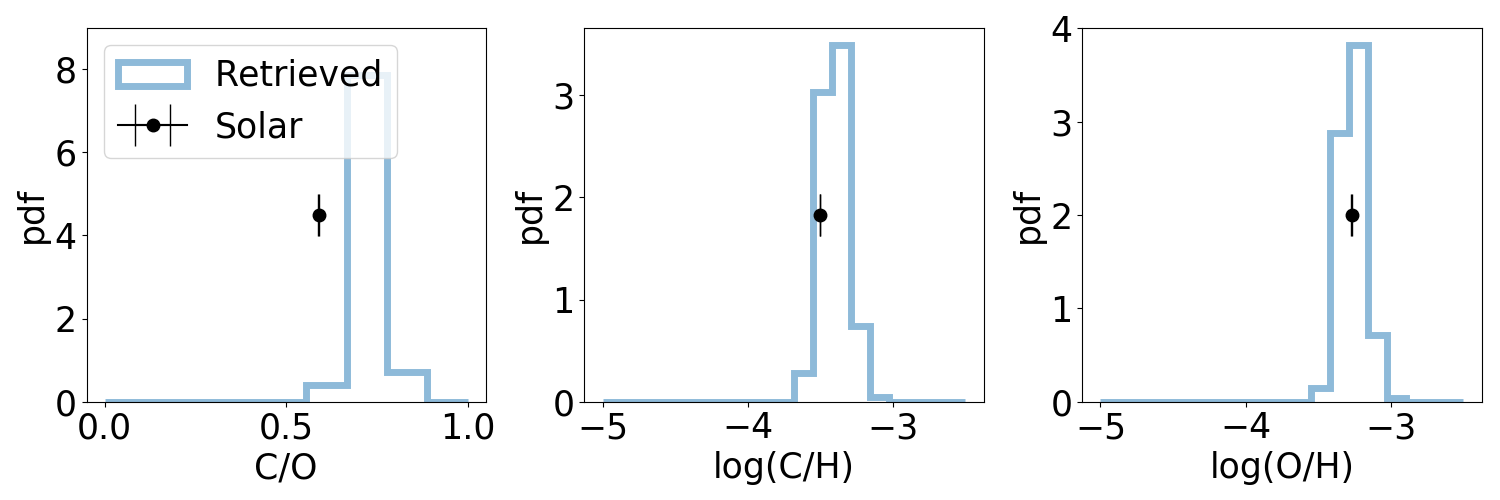} \\
\includegraphics[width=8.0cm]{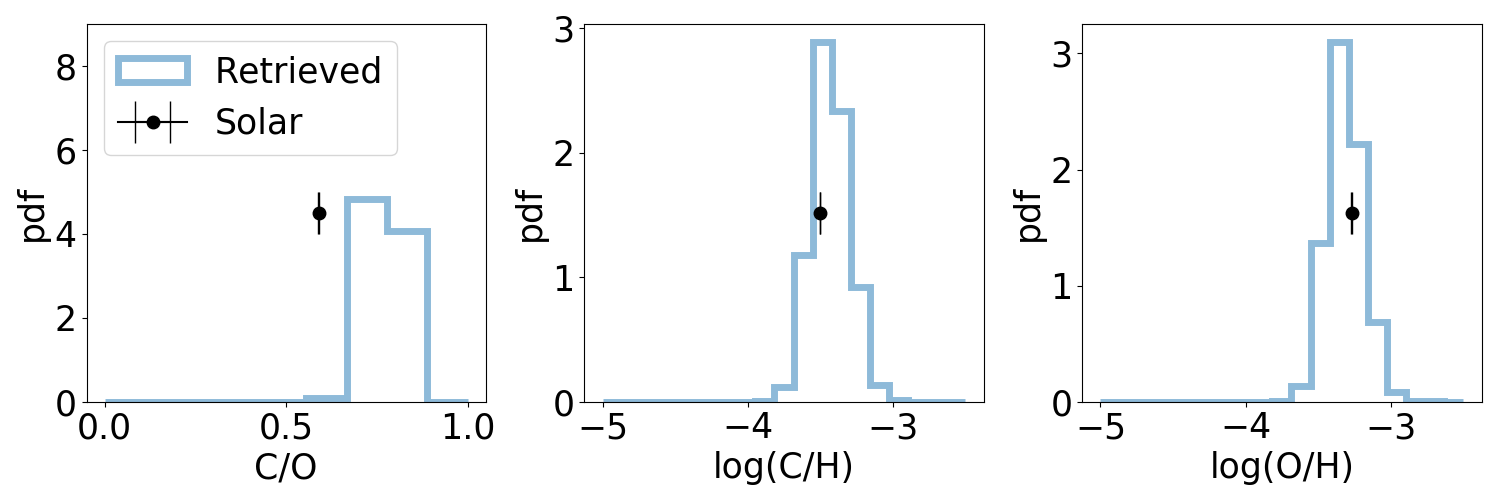}
\end{tabular}
\caption{{\bf{Retrieved C and O abundances and C/O based on the PHOENIX spectrum.}} While C and O abundances from posterior samples (blue histograms) agree well with the solar values (black data points), the retrieved C/O is consistently higher than the solar C/O. {\bf{Top}}: With a Gaussian prior (72.7$\pm$0.8 M$_{\rm{Jupiter}}$). {\bf{Bottom}}: With a flat mass prior between 10 and 100 M$_{\rm{Jupiter}}$. 
\label{fig:co_comp_phoenix}}
\end{figure}

\subsection{Retrieval Results}

\subsubsection{Fixed-Mass Case}

If the mass is tightly constrained, i.e., a benchmark BD as HR 7672 B with mass measured from the radial velocity technique and the astrometric data, we can apply a Gaussian mass prior. In the fixed-mass case for our PHOENIX retrieval, we apply a Gaussian prior of 72.7$\pm$0.8 M$_{\rm{Jupiter}}$ for mass and 5.5$\pm$0.2 for surface gravity. Our retrieval code can successfully reproduce the high-resolution spectral data and the photometric data (Fig. \ref{fig:data_model_phoenix_fixed}). A corner plot of the posterior distribution of all parameters is shown in Fig. \ref{fig:corner_fixed} and the 16th, 50th, and 84th percentiles of the posterior distributions are given in Table \ref{tab:mcmc_result}. 

Most importantly, the retrieved C and O abundance distributions encompass the solar values, which are used in the synthetic spectrum (Fig. \ref{fig:co_comp_phoenix}). However, the retrieved C/O is overestimated by 0.13 at 0.72$\pm$0.03 when compared to the solar value of 0.59. The disagreement is at 4-$\sigma$, implying potential systematics at the 0.15 level when retrieving C/O. We therefore adopt an uncertainty of 0.15 when reporting C/O values. The adopted uncertainty is also comparable with the C/O uncertainty of the solar C/O at 0.13 (Table \ref{tab:CO_abundances}).   

\subsubsection{Free-Mass Case}
\label{sec:free_mass}

We now consider a case with looser priors on mass and surface gravity. Instead of Gaussian priors, we apply a flat prior for mass and surface gravity, i.e., mass between 10 and 100 M$_{\rm{Jupiter}}$ and $\log$(g) between 3.5 and 5.5. The upper limit of 5.5 corresponds to the maximum surface gravity for a BD with a contraction time of the age of the universe. We recommend this free-mass prior be applied to most directly-imaged planets and BDs for which we do not have a tight mass and surface gravity constraint. In comparison, the informed Gaussian prior in the previous section is recommended for the tests on synthetic spectra and benchmark BDs for which surface gravity is well-constrained. 

\begin{figure}[h]
\begin{tabular}{l}
\includegraphics[width=8.0cm]{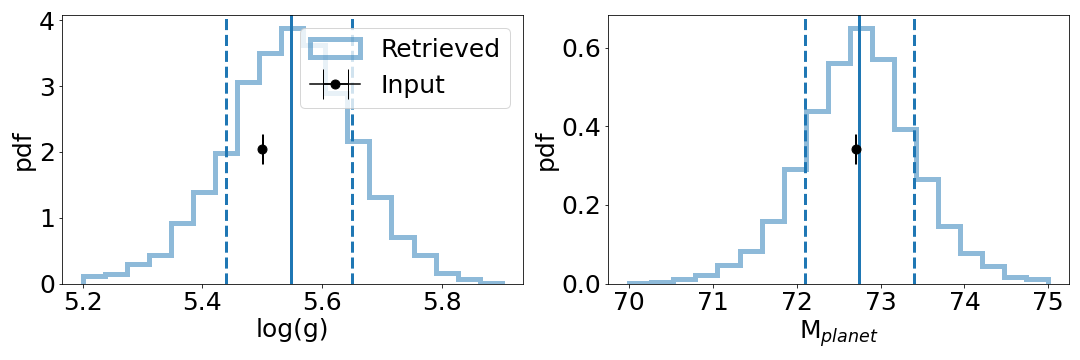} \\
\includegraphics[width=8.0cm]{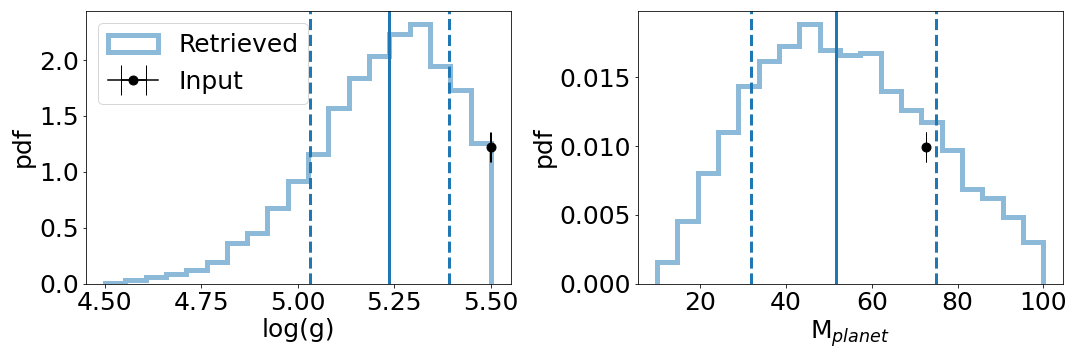}
\end{tabular}
\caption{{\bf{Retrieved surface gravity and mass based on the PHOENIX spectrum.}} Comparing surface gravity and mass from posterior samples (blue histograms) to PHOENIX input values (black data points) shows an agreement within 1-$\sigma$ for {\bf{Top}}: with a Gaussian prior (72.7$\pm$0.8 M$_{\rm{Jupiter}}$), and 1-2 $\sigma$ agreement for {\bf{Bottom}}: with a flat mass prior between 10 and 100 M$_{\rm{Jupiter}}$. Vertical lines mark the 16th, 50th, and 84th percentiles of posterior samples. 
\label{fig:logg_comp_phoenix}}
\end{figure} 

\begin{figure*}[ht!]
\epsscale{1.1}
\plotone{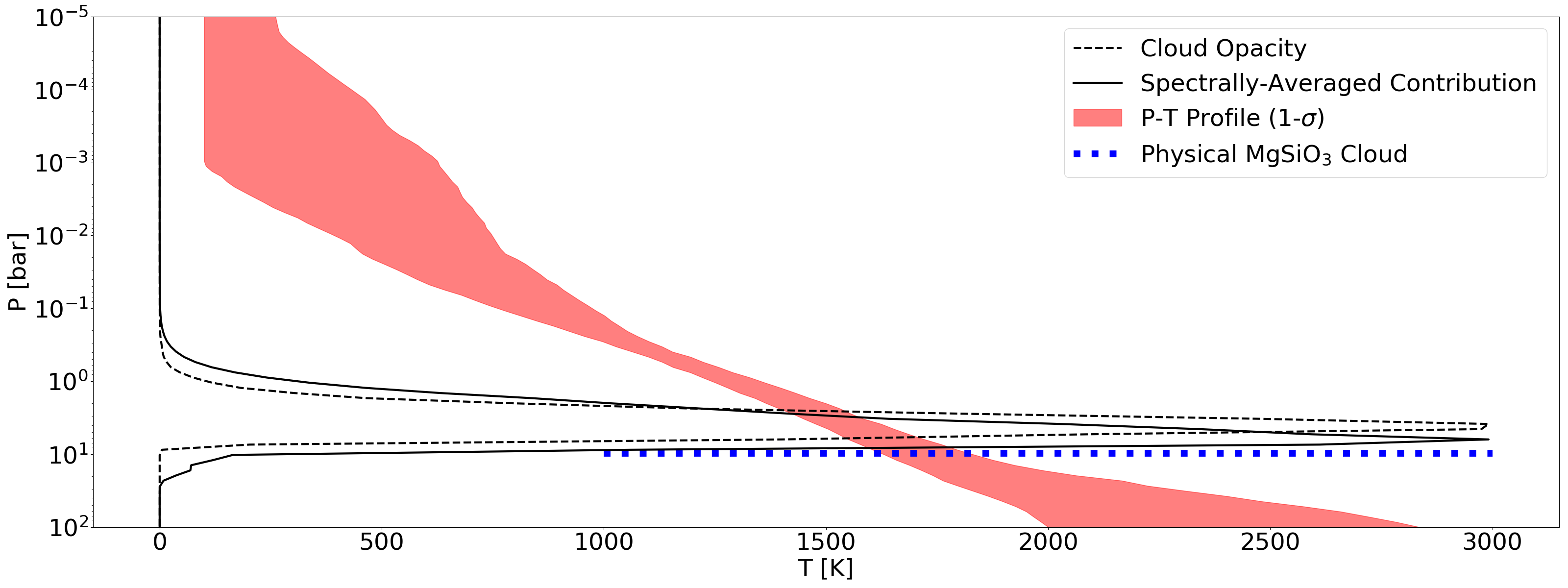}
\caption{{\bf{Retrieved P-T profile (1-$\sigma$ region in red shaded region) based on the PHOENIX spectrum.}} The spectrally-averaged contribution function is shown as the black solid line, which overlaps with the retrieved cloud layer (black dashed line). The cloud opacity does not significantly contribute to the emission because the optical depth of the cloud is small (\S \ref{sec:free_mass}). The pressure level of the retrieved cloud layer is consistent with that of a physical MgSiO$_3$ cloud (blue dotted line).  
\label{fig:pt_phoenix}}
\end{figure*} 

While we can successfully reproduce the high-resolution spectral data and the photometric data (Fig. \ref{fig:data_model_phoenix_free}), the retrieved $\log$(g) and mass are lower than input values by $\sim$1-2 $\sigma$ as shown in Table \ref{tab:mcmc_result} and Fig. \ref{fig:logg_comp_phoenix}. A complete corner plot is shown in Fig. \ref{fig:corner_free}. This may indicate that the high-resolution spectroscopic data and the photometric data points alone do not put a tight constraint on mass and surface gravity. 

The retrieved C and O abundances agree well with the solar values as shown in Fig. \ref{fig:co_comp_phoenix}. The retrieved C/O is overestimated by 0.18 at 0.77$\pm$0.04 when compared to the solar value of 0.59. This is again a $\sim$4-$\sigma$ discrepancy given the formal error bar from the retrieval analysis. If using the 0.15 adopted C/O uncertainty as discussed in the previous section, this is a 1.2-$\sigma$ discrepancy. 

Fig. \ref{fig:pt_phoenix} shows a number of pressure-dependent properties. First, the 1-$\sigma$ uncertainty region of the retrieved P-T profile is marked in red. When intercepting the condensation temperature of MgSiO$_3$~\citep[$\sim$1700 K, ][]{Marley2013}, the blue dotted line indicates the pressure level where MgSiO$_3$ clouds form. This is consistent (by design, see \S \ref{sec:cloud}) with the retrieved properties of the MgSiO$_3$ cloud whose opacity distribution is shown as the black dashed line. 

The contribution function (black solid line in Fig. \ref{fig:pt_phoenix}) coincides with the cloud opacity, indicating a cloudy condition. However, a closer look at the cloud opacity reveals that the cloud contributes negligible optical depth. The retrieved cloud opacity is $\sim$10$^{-6}$ g/cm$^2$. At a pressure of 3 bar and a temperature of 1500 K, where the cloud opacity peaks, the density is 5.5$\times10^{-5}$ g/cm$^3$ assuming an ideal gas law with a mean molecular weight of 2.3. Based on $\tau=\kappa\rho l$, where $\tau$ is optical depth, $\kappa$ is opacity, and $l$ is the atmosphere thickness, the optical depth is 0.005 even if we assume a thickness of 1000 km.  

In comparison, the cloudy condition is predicted by the BT-settl model~\citep[e.g., Fig. 4 in ][]{Allard2012} at an effective temperature of 1600 K. The discrepancy can be reconciled by the higher surface gravity (by 0.5 dex) that we consider here. At a higher surface gravity, the cloud deck sinks and therefore reveals a cloudless condition. 












\subsection{Potential Reasons for Overestimating C/O}
\label{sec:higherCO}

There may be a few caveats that lead to biases in estimating C/O. First, spectral normalization can affect C and O detection and retrieved abundances~\citep{Rasmussen2021}. In particular, in the presence of noise, the CO bandhead strength is likely to be underestimated as a result of spectral normalization. However, a reduced CO bandhead strength will lead to a lower C abundance, and therefore a lower C/O, which is the opposite to our result. 

Second, P-T profile can affect C/O measurements. As shown in~\citet{Wang2020}, a difference in parameterizing P-T profile leads to significantly different C/O measurements. This motivates the more flexible P-T profile used in this work, so we conclude that P-T profile is less likely to cause the C/O overestimation. 

Third, line saturation can bias C and O measurements. In $K$-band spectroscopy, CO lines tend to be deeper and H$_2$O tend to be weaker (Fig. \ref{fig:dataset}). Deep vs. saturated CO lines are less distinguishable in high-resolution retrieval after continuum normalization. This can lead to an overestimation of C abundance, which is consistent with our findings of the retrieval on the synthetic spectrum. A similar result is also found in~\citet{Finnerty2021} where overestimation of C/O is reported in retrievals for spectra with C/O lower than 0.5, although the bias in measuring C/O is smaller than 0.1.  

Lastly, weak H$_2$O lines tend to be interpreted as noise in the spectrum. This leads to an underestimation of O abundance and therefore a bias for a higher C/O, which is another plausible explanation for the higher C/O than solar value that is retrieved for the synthetic spectrum. 

In conclusion, the above exercise with a PHOENIX synthetic spectrum (1) tests the limits (e.g., retrieving mass and surface gravity) and estimates a more practical error bar (e.g., measuring C/O) and (2) validates our framework so that we can use it to retrieve C and O abundances based on the actual HR 7672 B data set.




\begin{figure*}[h]
\begin{tabular}{l}
\includegraphics[width=17.0cm]{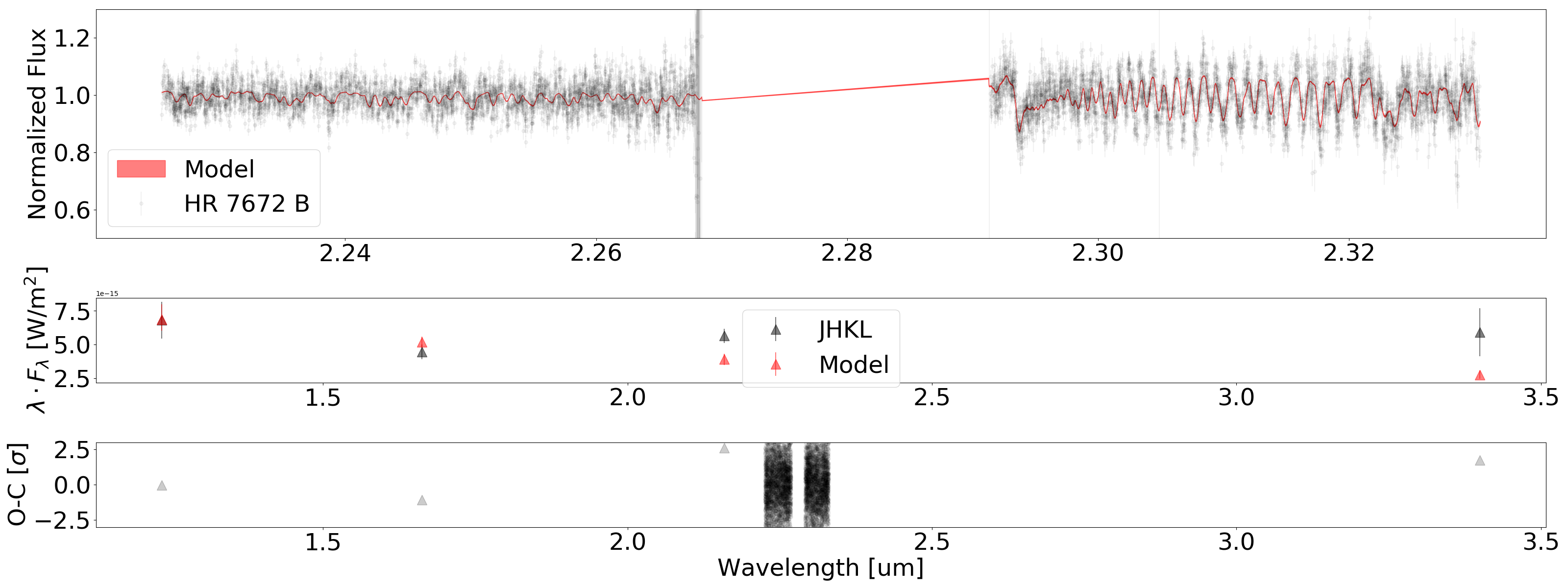} 
\end{tabular}
\caption{{\bf{Retrieved spectra for HR 7672 B with a Gaussian prior (72.7$\pm$0.8 M$_{\rm{Jupiter}}$.}}. Top two panels are simulated high-resolution spectroscopic and photometric data (black) and the modeled data (red). The bottom panel is a residual plot with data minus model and divided by errors. 
\label{fig:data_model_hr_fixed}}
\end{figure*} 

\begin{figure*}[h]
\begin{tabular}{l}
\includegraphics[width=17.0cm]{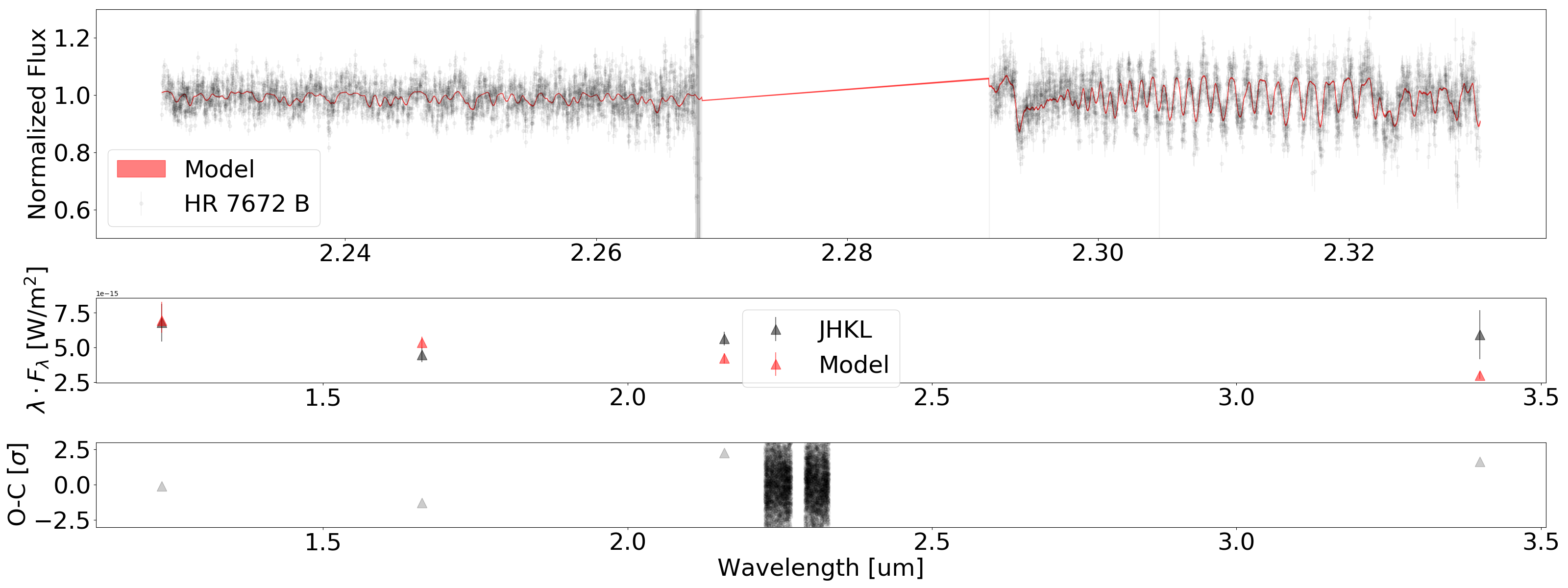}
\end{tabular}
\caption{{\bf{Retrieved spectra for HR 7672 B with a flat mass prior between 10 and 100 M$_{\rm{Jupiter}}$.}} Top two panels are simulated high-resolution spectroscopic and photometric data (black) and the modeled data (red). The bottom panel is a residual plot with data minus model and divided by errors. 
\label{fig:data_model_hr_free}}
\end{figure*} 

\section{Retrieving Properties for HR 7672 B}
\label{sec:combined_retrieval}

\subsection{Data vs. Modeled Spectra}

We use our retrieval framework to infer atmospheric C and O abundance for HR 7672 B using a data set that combines high-resolution data from KPIC and photometric data as detailed in \S \ref{sec:observation}. The retrieval setup as the same as our tests using the PHOENIX synthetic spectrum, except for one additional free parameter that accounts for the normalized flux offset between the high-resolution data and the modeled spectrum. This free parameter is introduced because of the uncertainty of normalizing the observed spectrum in the presence of noises.  


Similar to \S \ref{sec:test}, we consider two cases: the fixed-mass case and the free-mass case. The fixed-mass case is for retrieving objects with tight mass constraints. The free-mass case is for the majority of directly-imaged planets and BDs without tight mass constraints. Fig. \ref{fig:data_model_hr_fixed} and Fig. \ref{fig:data_model_hr_free} show the  comparison between data and models using posterior samples (Fig. \ref{fig:corner_hr_fixed} and Fig. \ref{fig:corner_hr_free}). In both cases, spectral models from posterior samples agree well with the observed spectrum and photometric data points except for the $K$- and $L$-band photometry, which shows $\sim$2-3 $\sigma$ discrepancy. The discrepancy can be attributed to a degeneracy in retrieving clouds and is discussed in \S \ref{sec:cloud_hr} and \S \ref{sec:dimmerKL}. 

{Despite the $\sim$2-3$\sigma$ discrepancy for the $K$- and $L$-band photometry, the retrieved luminosity for HR 7672 B ($\log{(L_{bol}/L_{\odot})}=-4.08\pm0.06$) agrees with literature values within $\sim$1-$\sigma$, e.g., $\log{(L_{bol}/L_{\odot})}=-4.19\pm0.04$~\citep{Brandt2019} and $\log{(L_{bol}/L_{\odot})}=-4.12\pm0.09$~\citep{Liu2002}. The retrieved effective temperature ($1806\pm77$ K) also falls in the previously estimated range between 1510 and 1850 K~\citep{Liu2002}.   

}


We also note that HR 7672 B is a very fast rotator with a v$\sin i$ of $45.0\pm0.5$ km$\cdot$s$^{-1}$, potentially making it an excellent object for Doppler imaging. Our v$\sin i$ is 1.8-$\sigma$ off the measurement by~\citet{Delorme2021} at $42.6\pm0.8$ km$\cdot$s$^{-1}$.

\subsection{Fixed Mass vs. Free Mass}

The mass of HR 7672 B is well-constrained because of available radial velocity data and astrometric data~\citep{Crepp2012, Brandt2019}. This corresponds to the fixed-mass case. However, we would like to investigate a case in which tight constraints are not available. This corresponds to the free-mass case, in which we set a flat prior for mass and surface gravity (see Table \ref{tab:prior}). 

\begin{figure}[h]
\begin{tabular}{l}
\includegraphics[width=8.0cm]{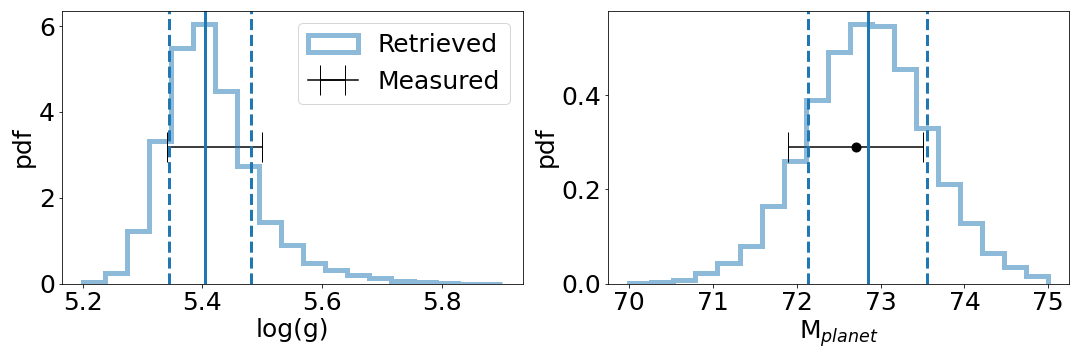} \\
\includegraphics[width=8.0cm]{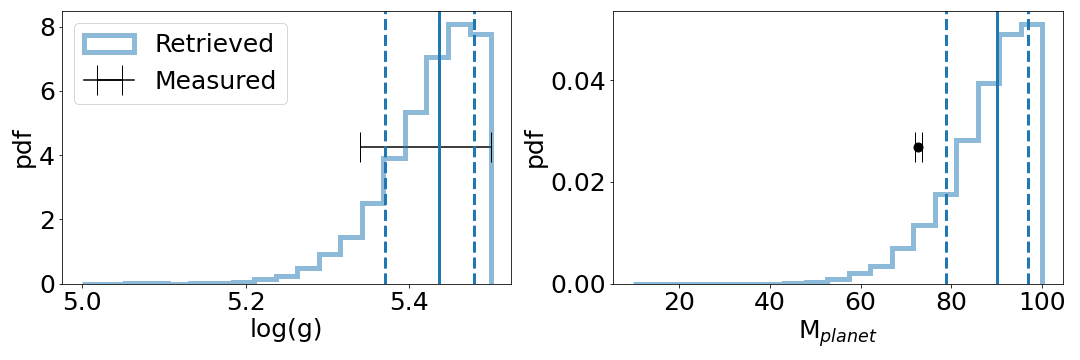}
\end{tabular}
\caption{{\bf{Retrieved surface gravity and mass for HR 7672 B.}} {\bf{Top}} with a Gaussian prior (72.7$\pm$0.8 M$_{\rm{Jupiter}}$): retrieved surface gravity and mass (blue histograms) agree well with measured values (black data points with error bars). {\bf{Bottom}} with a flat mass prior between 10 and 100 M$_{\rm{Jupiter}}$: while the retrieved surface gravity agrees with the measured value, the mass posteriors differ by 1-2 $\sigma$. Vertical lines mark the 16th, 50th, and 84th percentiles of posterior samples.  
\label{fig:logg_comp_hr}}
\end{figure}

For the fixed-mass case, the retrieved surface gravity and mass posteriors agree well with measured values (Fig. \ref{fig:logg_comp_hr} Top and Table \ref{tab:mcmc_result}). The mass constraints are from radial velocity and astrometric data~\citep{Brandt2019}. The surface gravity constraints are from the age measurement of $\sim$2-4 Gyr~\citep{Brandt2019} and BD evolutionary models~\citep{Saumon2008}. Assuming the HR 7672 system has an age of between 2 and 4 Gyr~\citep{Brandt2019}, the $\log$(g) of HR  7672 B should be between 5.33 and 5.35 based on~\citet{Saumon2008} and surely lower than 5.50. For the free-mass case (Fig. \ref{fig:logg_comp_hr} Bottom), while the retrieved surface gravity agrees with the ~\citep{Saumon2008} measured value, the mass posteriors differ by 1-2 $\sigma$. The 1-2 $\sigma$ difference is consistent with our findings using the PHOENIX synthetic spectrum.

The comparison between the fixed-mass case and the free-mass case shows that our retrieval framework can retrieve mass and surface gravity within 1-2 $\sigma$ for BDs like HR 7672 B. Further tests are needed for directly-imaged exoplanets with lower surface gravity.

\subsection{C/O}

\begin{figure}[h]
\begin{tabular}{l}
\includegraphics[width=8.0cm]{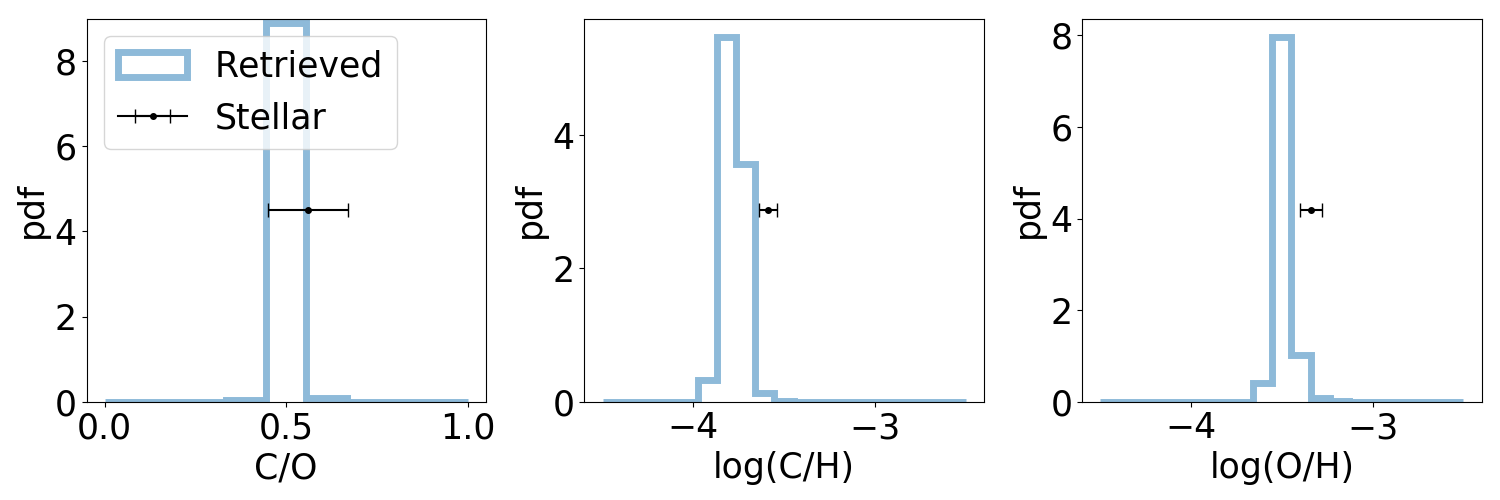} \\
\includegraphics[width=8.0cm]{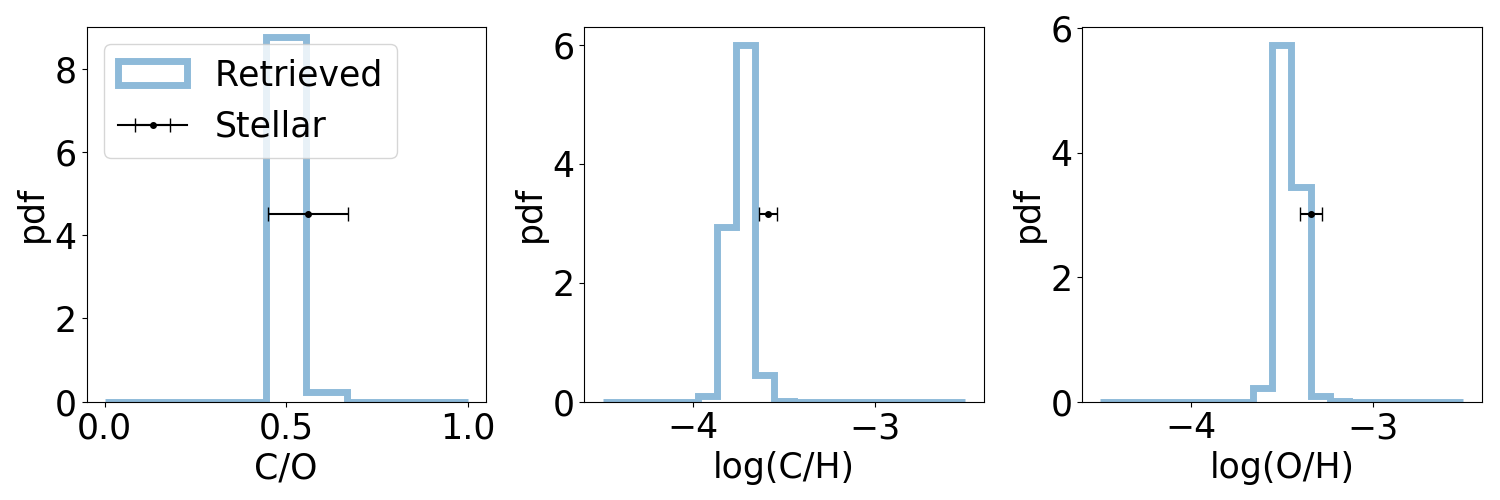}
\end{tabular}
\caption{{\bf{Retrieved C and O abundances and C/O for HR 7672 B.}} Comparing C/O, C/H, and O/H from posterior samples (blue histograms) to stellar values (black data points with error bars) shows an agreement within 1-2 $\sigma$. {\bf{Top}}: With a Gaussian prior (72.7$\pm$0.8 M$_{\rm{Jupiter}}$). {\bf{Bottom}}: With a flat mass prior between 10 and 100 M$_{\rm{Jupiter}}$. 
\label{fig:co_comp_hr}}
\end{figure}

Moreover, the retrieved C and O abundances and C/O for HR 7672 B agree within 1-2 $\sigma$ to those of HR 7672 A (Fig. \ref{fig:co_comp_hr}). The consistency is seen for both the fixed-mass case and the free-mass case. Using the free-mass case as an example, the retrieved [C/H], [O/H], and C/O are below stellar values by 1.5-$\sigma$, 1.2-$\sigma$, and 0.3-$\sigma$, respectively. From previous work on benchmark BDs, $<$2- $\sigma$ difference is considered in good agreement~\citep{Line2015}. 

{We also note that the retrieved uncertainties for [C/H], [O/H], and C/O are likely underestimated. This is not uncommon in recent papers that perform retrieval analyses on directly-image exoplanets and brown dwarfs. For example, reported C/O uncertainty is 0.06-0.07 in ~\citet{Molliere2020} and ~\citet{Burningham2021}, much lower than the solar C/O uncertainty at 0.13~\citep{Palme2014}. Additional unaccounted for systematic errors can exceed the formal uncertainties. This is evidenced by the C/O discrepancy as seen in \S \ref{sec:free_mass} for the PHOENIX retrieval and possible explanations are discussed in \S \ref{sec:higherCO}. Small wavelength coverage in high spectral resolution data may also contribute to the additional unknown systematics, which can be mitigated by increased spectral grasp.

}

In addition, there is a subtle difference between atmospheric abundance and intrinsic abundance as pointed out by~\citet{Line2015}. The atmospheric abundance is retrieved based on the BD spectrum and the intrinsic abundance is measured based on the primary star. The latter is intrinsic because of efficient mixing in the photosphere of the primary star. The atmospheric abundance from a BD can be affected by condensation, which will decrease oxygen abundance because a condensed particle is likely to contain oxygen, e.g., MgSiO$_3$. However, our retrieved MgSiO$_3$ abundance is at lower than 10$^{-4}$ level. Therefore, the oxygen locked in MgSiO$_3$ will not significantly affect the retrieved O/H and C/O when comparing the MgSiO$_3$ abundance to the major O carrier H$_2$O and CO for which the abundance is at 10$^{-3}$-10$^{-2}$ level. 



\begin{figure*}[ht]
\epsscale{1.1}
\plotone{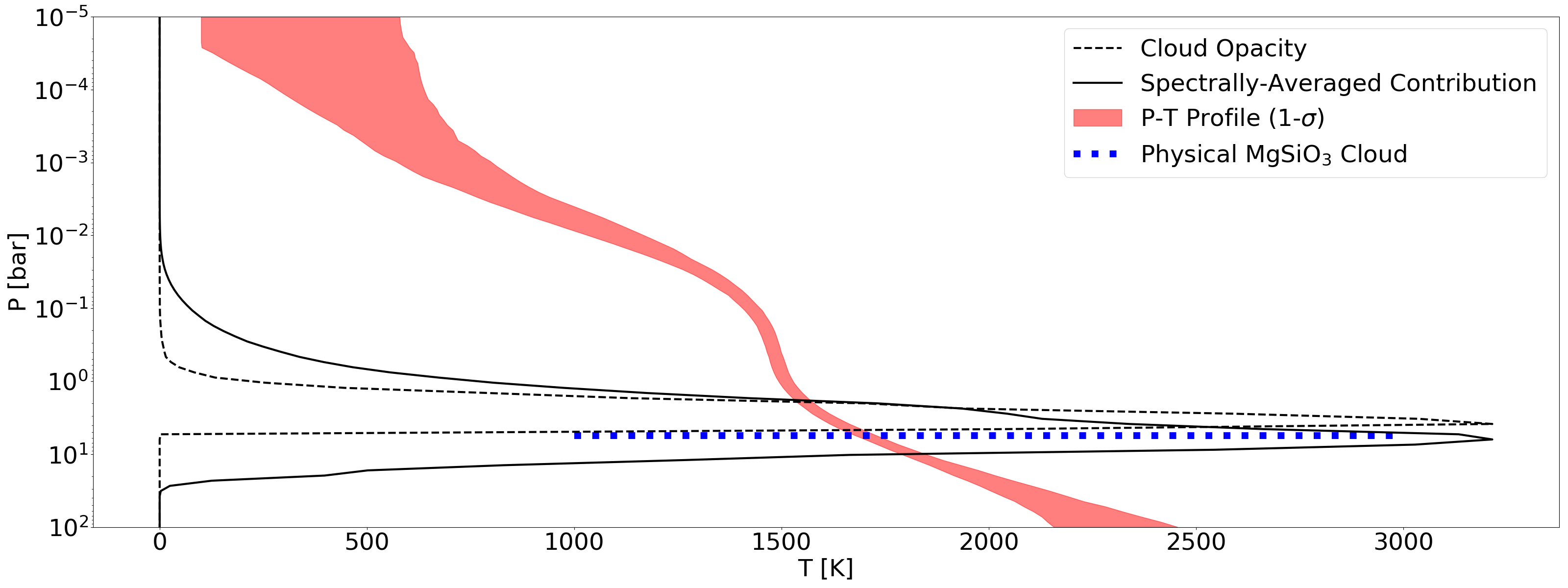}
\caption{{\bf{Retrieved P-T profile (1-$\sigma$ region in red shaded region) for HR 7672 B.}}. The spectrally-averaged contribution function is shown as the black solid line, which overlaps with retrieved cloud layer (black dashed line). The cloud opacity does not significantly contribute to the emission because the optical depth of the cloud is small (see \S \ref{sec:cloud_hr}). The pressure level of the retrieved cloud layer is consistent with that of a physical MgSiO$_3$ cloud (blue dotted line). The isothermal knee of the P-T profile between 0.1 and 1 bar may be responsible for the cloudless inference. The degeneracy between a cloudy atmosphere and a cloudless atmosphere with a reduced thermal gradient (e.g., the isothermal knee) in the P-T profile is discussed in \S \ref{sec:cloud_hr}. 
\label{fig:pt_hr}}
\end{figure*} 

\begin{figure*}[h!]
\epsscale{1.1}
\plotone{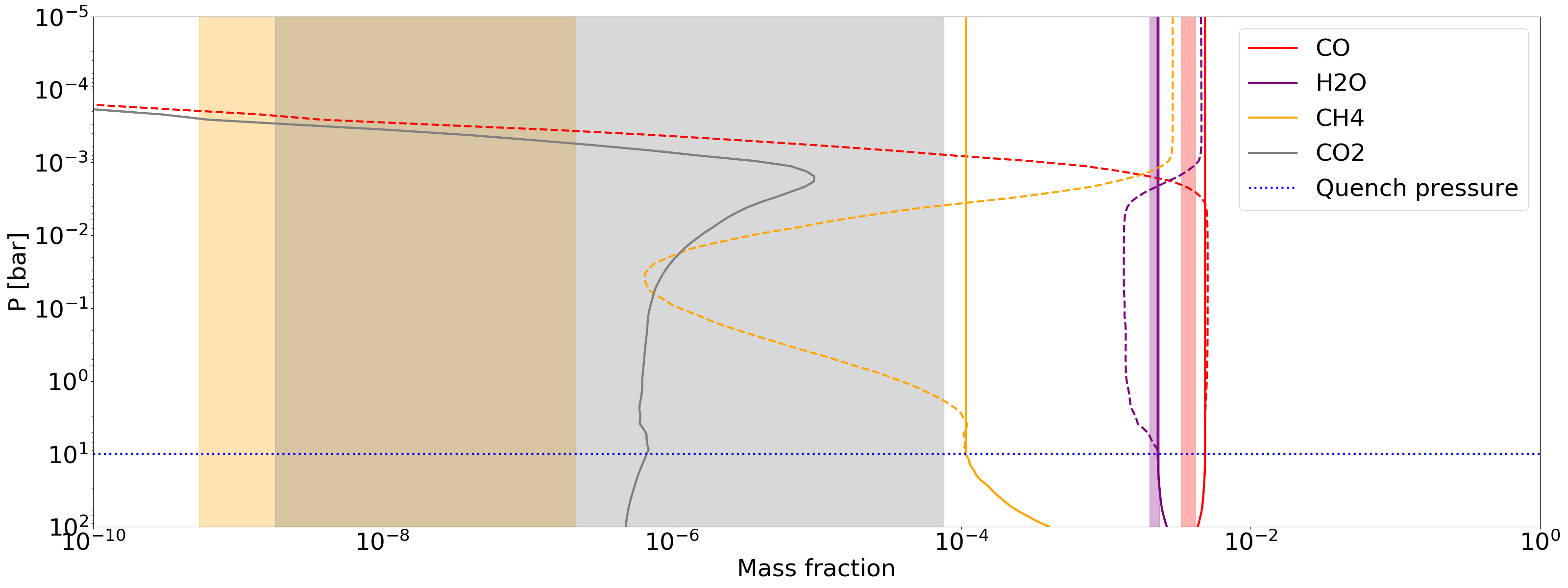}
\caption{Abundances assuming equilibrium chemistry (solid) and a quenched pressure at 10 bar (dashed). 1-$\sigma$ ranges of retrieved abundances are shown in shaded regions. The expected CO and H$_2$O abundances are much higher than those of CO$_2$ and CH$_4$, so CO and H$_2$O are two dominant C and O carriers. This is consistent with detected CO and H$_2$O lines in $K$-band spectroscopy (Fig. \ref{fig:dataset}. )
\label{fig:chem_equi}}
\end{figure*} 

\subsection{Cloud Property and Degeneracy}
\label{sec:cloud_hr}

Fig. \ref{fig:pt_hr} shows the retrieved P-T profile as well as the contribution function. Similarly to the PHOENIX retrieval case, we infer a cloudless condition even though the cloud opacity peaks roughly near the peak of the contribution function. This is because the optical depth due to the cloud is negligible given the low opacity value that is smaller than $10^{-5}$ g/cm$^2$. Moreover, cloud properties are mostly unconstrained as shown in Fig. \ref{fig:corner_hr_fixed} and Fig. \ref{fig:corner_hr_free}. 

The inferred cloudless condition may be due to a degeneracy as discussed in~\citet{Tremblin2017}: the existence of clouds can be masqueraded by a decrease of thermal gradient of the P-T profile. Both clouds and a nearly isothermal P-T profile can lead to shallower absorption lines. The isothermal knee around 1 bar in our retrieved P-T profile is the evidence of this degeneracy. Similar effects are also discussed in~\citet{Molliere2020}. Therefore, we do not know if the inferred cloudless condition is real or due to an artificially isothermal P-T profile. 

One solution to break the degeneracy is to use a self-consistent P-T profile as done in~\citet{Molliere2020}. However, a self-consistent P-T profile may not necessarily be the actual P-T profile in the BD atmosphere. Therefore, future JWST data will play a key role in resolving this degeneracy~\citep{Tremblin2017}. 

\subsection{The Fainter Retrieved $K$- and $L$-band Photometry Than Observation}
\label{sec:dimmerKL}

The degeneracy under discussion here may also explain the 2-3 $\sigma$ discrepancy of $K$- and $L$-band photometry that is seen in the retrieval for HR 7672 B (see Fig. \ref{fig:data_model_hr_fixed} and Fig. \ref{fig:data_model_hr_free}). While the \texttt{MultiNest} sampling algorithm preferentially finds the cloudless solution for the reason that is detailed in the next paragraph, an alternative cloudy solution can tilt the spectral energy distribution of the $J$, $H$, $K$, and $L$ bands, namely, making $J$ and $H$-band dimmer while leaving $K$ and $L$-band relatively unchanged. This is because clouds affect shorter wavelengths more than longer wavelengths. In this way, all four modeled photmetric data points are consistently below the actual measurement points. Thus, to account for the overall fainter modeled photometry than the observation, the retrieved radius is inflated in the retrieval and therefore results in a lower surface gravity. 

The reason why the retrieval favors the cloudless solution is that the weight for photometric data points is small: there are only 4 photometric data points whereas there are over 7000 spectroscopic data points. It is expected that when the weight of photometric data points increases the discrepancy of photometry is reduced. Indeed, this is what happens when increasing the photmetric weight by adding redundant photometric data points that repeat themselves. However, the retrieved surface gravity becomes too low to be realistic. This delicate issue will be discussed in more detail in a forthcoming paper.

\subsection{Chemical Equilibrium?}
\label{sec:chemical_eq}

We investigate the agreement between our retrieved abundances and those expected from chemical equilibrium. The comparison helps to check if reasonable abundances are retrieved. 

We use \texttt{poor\_mans\_nonequ\_chem} to interpolate a pre-calculated chemical grid from \texttt{easyCHEM}~\citep{Molliere2017}. The grid spans multiple dimensions including temperature (60 - 4000 K), pressure (10$^{-8}$ - 1000 bar), C/O (0.1 - 1.6) and [Fe/H] (-2 - 3). To calculate the equilibrium abundance, we use the median of the retrieved P-T profile and stellar values of C/O = 0.56 and [Fe/H] = -0.04. While the uncertainties of the P-T profile, C/O, and [Fe/H] all contribute to the uncertainty of the equilibrium abundance, we show below that the adopted values for the chemical grid result in reasonable agreement with the retrieved abundance.

Fig. \ref{fig:chem_equi} shows the abundances (in mass mixing ratio) for four species (CO, H$_2$O, CH$_4$, and CO$_2$) assuming two conditions: chemical equilibrium, and a quenched case of chemical disequilibrium, where vertical mixing homogenizes abundances above a quench pressure which we set at 10 bar.

For the two constrained species (CO and H$_2$O), the H$_2$O abundance agrees well with the quenched condition, but the retrieved CO abundance is below the value as interpolated from the chemical grid (by $\sim$1-$\sigma$). The difference can be reconciled by varying the quench pressure, the P-T profile, C/O, and [Fe/H] values that are within the posterior range. A more rigorous approach would be to sample the posterior and infer a range of possible values form the chemical grid. 

For the two unconstrained species (CH$_4$ and CO$_2$), the 16\% - 85\% credible range for CO$_2$ agrees well with both the chemical equilibrium and the quenched conditions. However, the retrieved CH$_4$ abundance range is well below the value as expected from the chemical grid.

\section{Summary}
\label{sec:summary}

This paper aims towards the goal of measuring chemical composition to better understand the origin of sub-stellar companions. First, we measure stellar abundance for HR 7672 A using archival data from the Keck Observatory Archive. The resulting stellar parameters and abundance for C and O are reported in Table \ref{table:Params} and Table \ref{tab:CO_abundances}, which are in $<$2-$\sigma$ agreement with previous measurements. 

Second, using KPIC, we obtain high-resolution (R=35,000) data for HR 7672 B, a benchmark BD around HR 7672 A. We measure $L$-band photometry for HR 7672 B using Keck NIRC2 archival data. Along with previous $J$, $H$, and $K$-band photometric data points, the spectrum and photometric data points are used to validate a retrieval framework, which is an extension of \petit. We show that the framework can retrieve correctly the C and O abundances that are used in a synthetic PHOENIX BT-Settle spectrum (\S \ref{sec:test}). However, the retrieved C/O is overestimated by 0.13-0.18 (4-$\sigma$) using the formal uncertainty from the retrieval. We therefore recommend a 0.15 uncertainty for the retrieved C/O. We also show that our retrieval framework can retrieve C and O abundances and C/O from a benchmark BD HR 7672 B that are within 1.5-$\sigma$ consistent with the primary star HR 7672 A (\S \ref{sec:combined_retrieval}). 

The work presented here provides a practical procedure of testing and performing atmospheric retrieval on data sets that span a large range of spectral resolution (e.g., from photometric data to R of 35,000) and wavelength coverage ($J$ through $L$-band). Our exercises on a synthetic spectrum and the HR 7672 B data set enable us to understand the limitations and uncertainties in retrieving BD properties and lend confidence in using the framework on future data sets from more BDs and exoplanets.\\

\noindent
{\bf{Acknowledgments}} We would like to thank Paul Molliere for the help in setting up and running \petit. We thank Anjali Piette for helpful discussion on P-T profile. We thank the Heising-Simons Foundation for supporting the workshop on combining high-resolution spectroscopy and high-contrast imaging for exoplanet characterization, where the idea originated on combining photometric data and spectral data of different resolutions. KPIC has been supported by the Heising-Simons Foundation through grants \#2015-129, \#2017-318 and \#2019-1312. This work was also partially supported by the Simons Foundation. The data presented herein were obtained at the W. M. Keck Observatory, which is operated as a scientific partnership among the California Institute of Technology, the University of California and the National Aeronautics and Space Administration. The Observatory was made possible by the generous financial support of the W. M. Keck Foundation. The authors wish to recognize and acknowledge the very significant cultural role and reverence that the summit of Mauna Kea has always had within the indigenous Hawaiian community.  We are most fortunate to have the opportunity to conduct observations from this mountain.  

%






\appendix
\section{Corner Plots for Retrievals}
\label{app:corner}

\begin{figure}[h!]
\epsscale{1.0}
\plotone{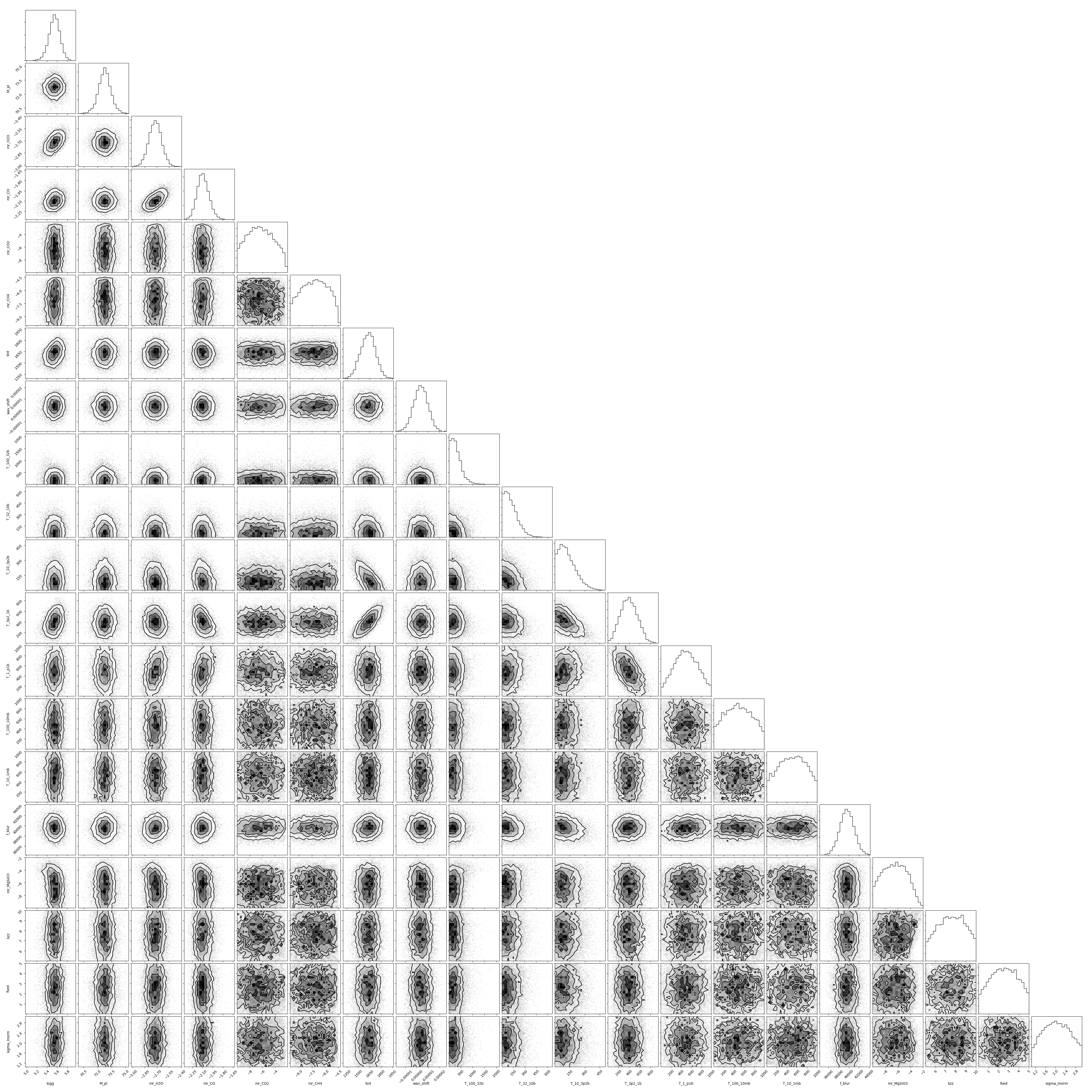}
\caption{Posterior distributions for the PHOENIX retrieval by fixing the mass to 72.7$\pm$0.8 M$_{Jupiter}$ (see \S \ref{sec:test}).
\label{fig:corner_fixed}}
\end{figure} 

\begin{figure}[h!]
\epsscale{1.0}
\plotone{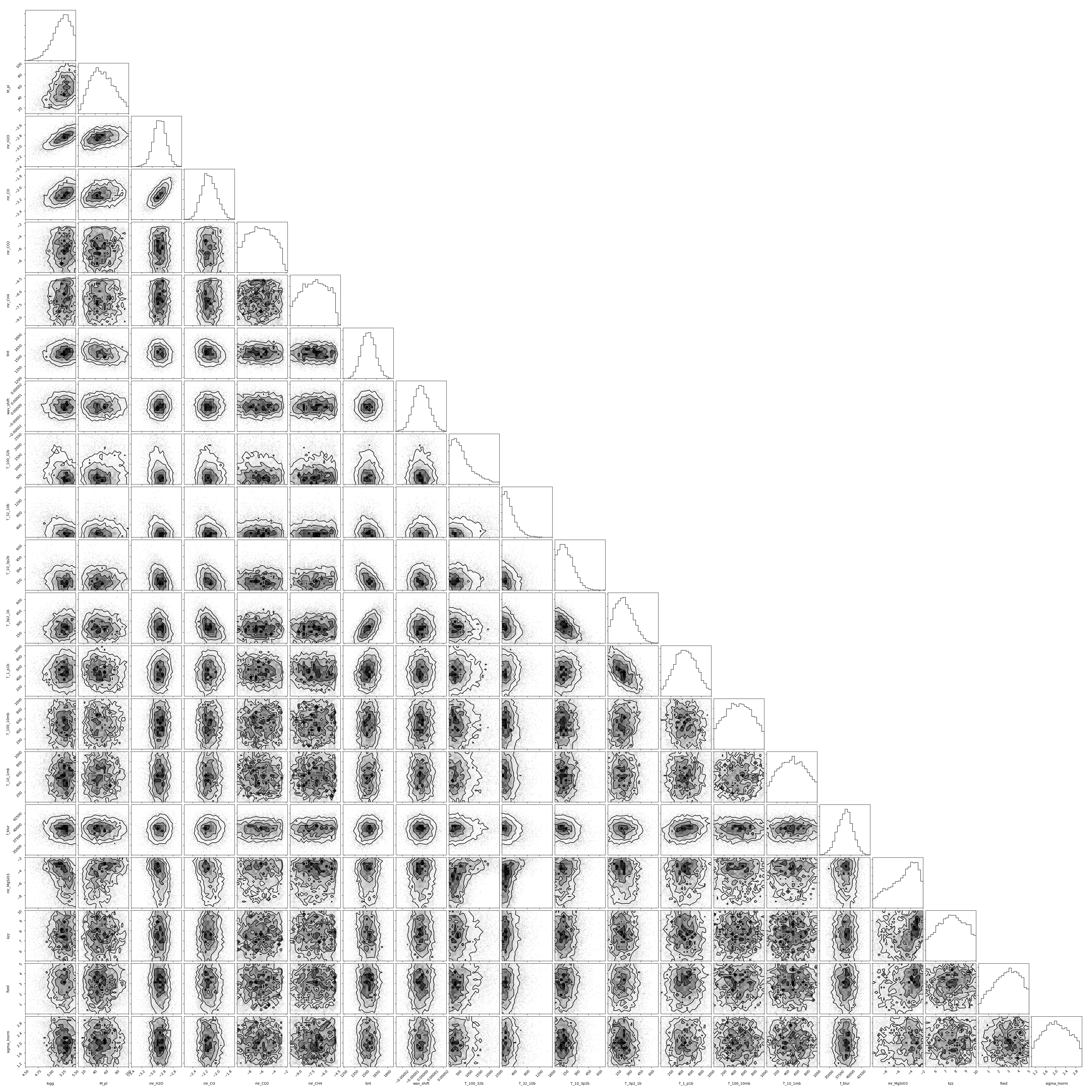}
\caption{Posterior distributions for the PHOENIX retrieval with a flat prior between 10 and 100 M$_{Jupiter}$ (see \S \ref{sec:test}).
\label{fig:corner_free}}
\end{figure} 

\begin{figure}[h!]
\epsscale{1.0}
\plotone{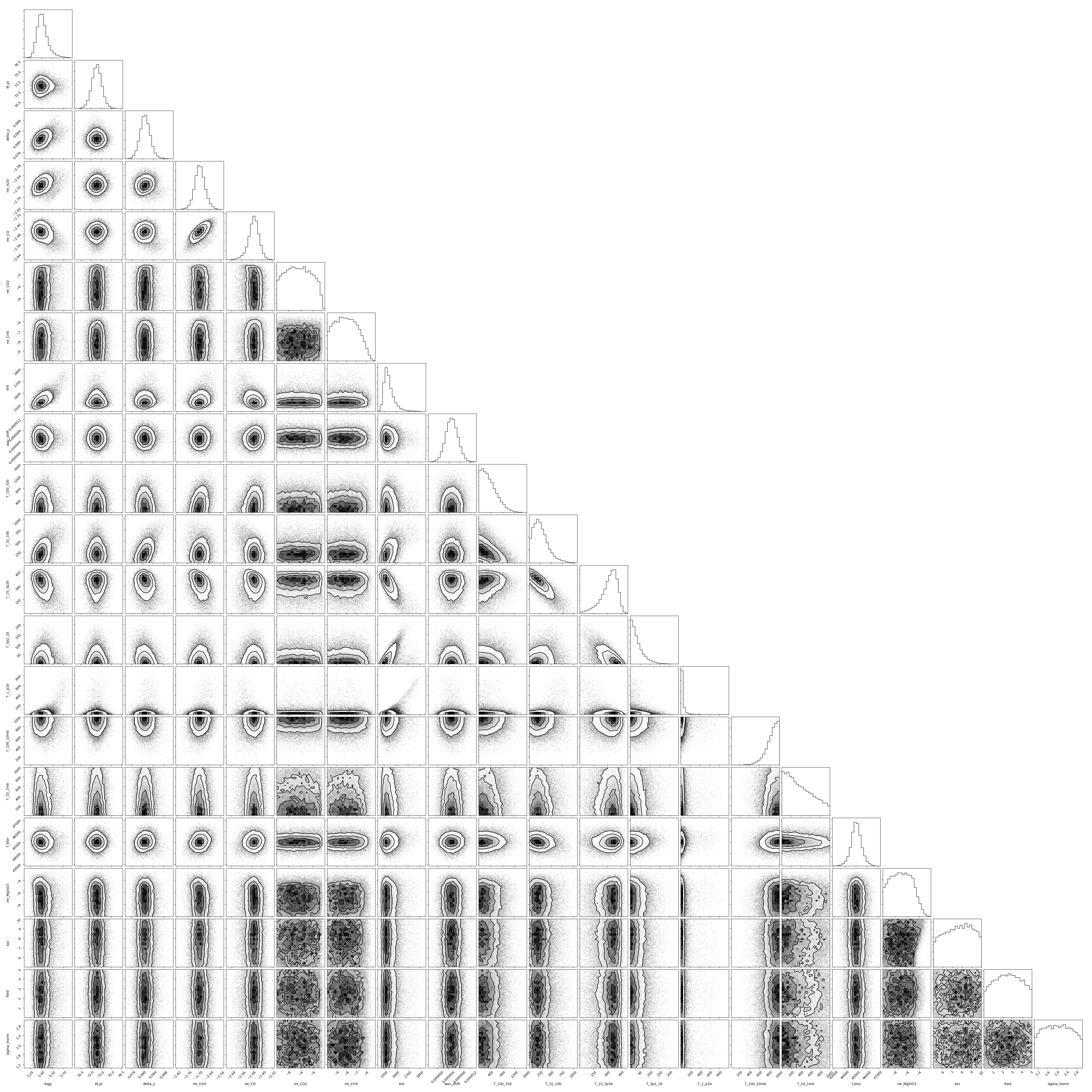}
\caption{Posterior distributions for the retrieval for HR 7672 B by fixing the mass to 72.7$\pm$0.8 M$_{Jupiter}$ (see \S \ref{sec:combined_retrieval}).
\label{fig:corner_hr_fixed}}
\end{figure} 

\begin{figure}[h!]
\epsscale{1.0}
\plotone{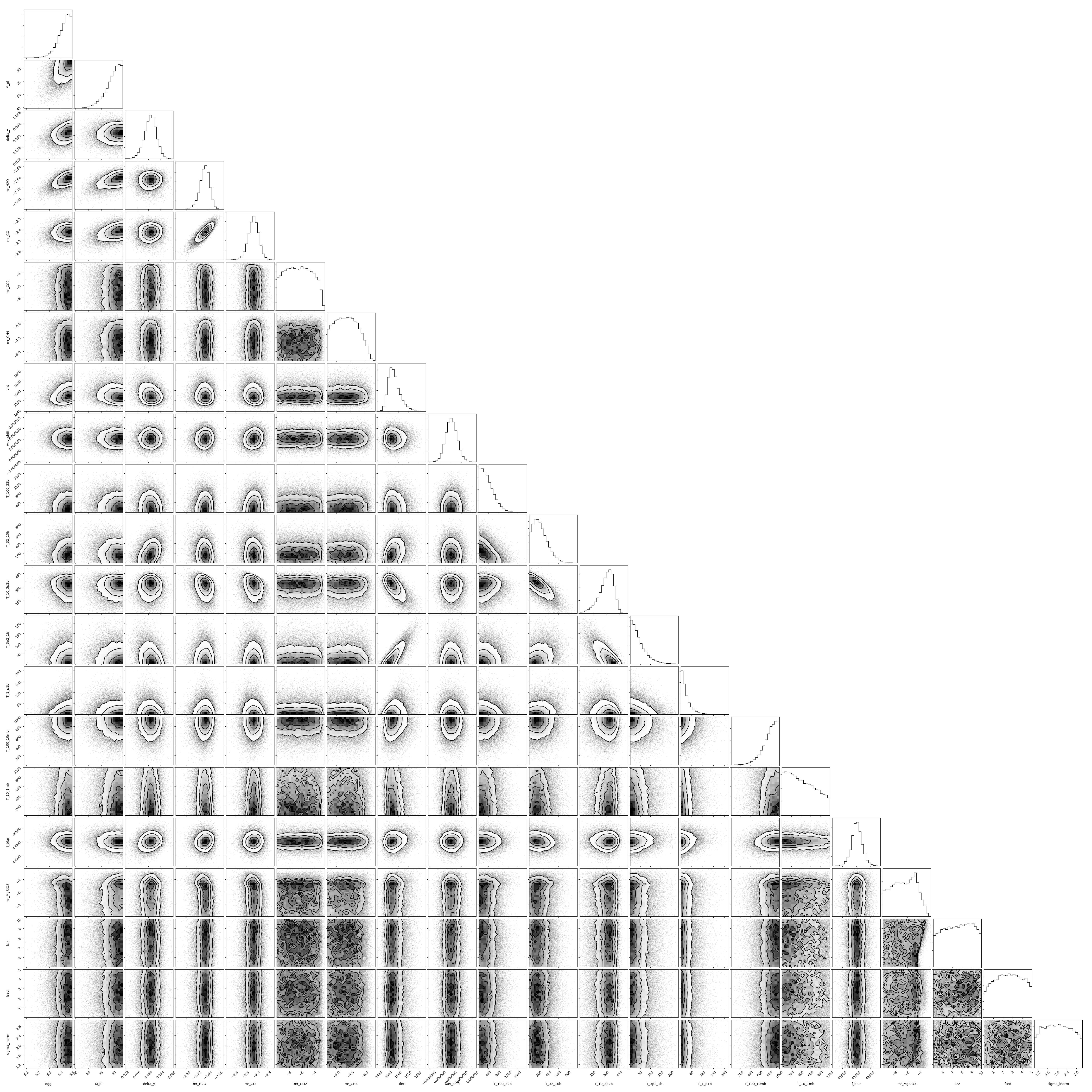}
\caption{Posterior distributions for the retrieval for HR 7672 B with a flat prior between 10 and 100 M$_{Jupiter}$ (see \S \ref{sec:combined_retrieval}).
\label{fig:corner_hr_free}}
\end{figure}

\bibliography{sample63}{}
\bibliographystyle{aasjournal}



\begin{table}
\scriptsize
\centering
 \begin{tabular}{| c | c | c | c | c |} 
 \hline
  & \teff &  $\log$(g) & $\xi$ (km/s) & [Fe/H]\\ [0.5ex] 
 \hline\hline
 This work & 5946$\pm$40 & 4.43$\pm$0.01 & 0.81$\pm$0.16 & -0.04$\pm$0.07\\
\citet{daSilva2015} & 5972$\pm$34 & 4.44$\pm$0.14 & 1.03$\pm$0.08 & 0.08$\pm$0.05\\
\citet{Brewer+2016} & 5940$\pm$25 & 4.40$\pm$0.03 & -- & 0.07$\pm$0.01\\
\citet{Luck2017} & 5946 & 4.40 & 1.21 & 0.04$\pm$0.16\\
 \hline
\end{tabular}
\caption{Stellar parameters derived by this paper and by \citet{daSilva2015}, \citet{Brewer+2016} and \citet{Luck2017}} \label{table:Params}
\end{table}




\begin{deluxetable}{clllll}
\tablewidth{0pt}
\tablecaption{C and O abundances for HR~7672 A and brown dwarf HR 7672 B (Note: Abundances relative to solar are calculated with respect to the solar values used by each individual paper, which differ from the solar reference used in this work~\citep{Palme2014}.) \label{tab:CO_abundances}}
\tablehead{
\colhead{\textbf{}} &
\colhead{\textbf{$\log{\epsilon_{\rm C}}$}} &
\colhead{\textbf{[C/H]}} &
\colhead{\textbf{$\log{\epsilon_{\rm O}}$}} &
\colhead{\textbf{[O/H]}} &
\colhead{\textbf{C/O}} 
}


\startdata
\multicolumn{6}{l}{HR 7672 A} \\
\hline
This work &8.41$\pm$0.05 & -0.09$\pm$0.05 & 8.66$\pm$0.06 & -0.07$\pm$0.06 & 0.56$\pm$0.11 \\
\citet{daSilva2015} & 8.51$\pm$0.03 & -0.05$\pm$0.03 & -- & -- & -- \\
\citet{Brewer+2016} & 8.45$\pm$0.03 & 0.06$\pm$0.03 & 8.72$\pm$0.04 & 0.06$\pm$0.04 & 0.53$\pm$0.06 \\
\citet{Luck2017} & 8.40$\pm$0.15 & -0.03$\pm$0.15 & 8.63$\pm$0.15 & -0.06$\pm$0.15 & 0.59$\pm$0.10 \\
Solar & 8.50$\pm$0.06 & 0.00$\pm$0.06 & 8.73$\pm$0.07 & 0.00$\pm$0.07 & 0.59$\pm$0.13 \\
\hline
\multicolumn{6}{l}{HR 7672 B} \\
\hline
Fixed-mass &8.23$^{+0.04}_{-0.04}$ & -0.27$^{+0.04}_{-0.04}$ & 8.51$^{+0.04}_{-0.03}$ & -0.22$^{+0.04}_{-0.03}$ & 0.52$^{+0.02}_{-0.02}$ \\
Free-mass &8.26$^{+0.05}_{-0.05}$ & -0.24$^{+0.05}_{-0.05}$ & 8.54$^{+0.04}_{-0.04}$ & -0.19$^{+0.04}_{-0.04}$ & 0.52$^{+0.02}_{-0.02}$ \\
\enddata

\end{deluxetable}

\begin{deluxetable}{lcccc}
\tablewidth{0pt}
\tablecaption{Parameters used in retrieval and their priors.\label{tab:prior}}
\tablehead{
\colhead{\textbf{Parameter}} &
\colhead{\textbf{Unit}} &
\colhead{\textbf{Type}} &
\colhead{\textbf{Lower}} &
\colhead{\textbf{Upper}} \\
\colhead{\textbf{}} &
\colhead{\textbf{}} &
\colhead{\textbf{}} &
\colhead{\textbf{or Mean}} &
\colhead{\textbf{or Std}} 
}

\startdata
Fixed Surface Gravity ($\log$(g))                   &  cgs                 &   Gaussian        &  5.5        &   0.2       \\
Free Surface Gravity ($\log$(g))                   &  cgs                 &   Uniform        &  3.5        &   5.5       \\
Fixed Mass (M$_P$)                       &  M$_{\rm{Jupiter}}$  &   Gaussian        &   72.7      &   0.8       \\
Free Mass (M$_P$)                       &  M$_{\rm{Jupiter}}$  &   Uniform        &   10      &   100       \\
H$_2$O Mixing Ratio ($\log$(mr$_{\rm{H}_2\rm{O}}$)) &  \nodata             &   Log-uniform      &  -10       &   -1      \\
CO Mixing Ratio ($\log$(mr$_{\rm{C}\rm{O}}$))       &  \nodata             &   Log-uniform       &   -10     &   -1      \\
CO$_2$ Mixing Ratio ($\log$(mr$_{\rm{C}\rm{O}_2}$)) &  \nodata             &   Log-uniform       &  -10       &   -1      \\
CH$_4$ Mixing Ratio ($\log$(mr$_{\rm{C}\rm{H}_4}$)) &  \nodata             &   Log-uniform       &  -10       &   -1      \\
Temperature at 3.2 bar (t$_{\rm{int}}$)      &  K                   &   Uniform   &   1000 &   2000  \\
$\Delta T$ between 100 and 32 bar      &  K                   &   Uniform   &   0 &   2500  \\
$\Delta T$ between 32 and 10 bar      &  K                   &   Uniform   &   0 &   2000  \\
$\Delta T$ between 10 and 3.2 bar      &  K                   &   Uniform   &   0 &   1500  \\
$\Delta T$ between 3.2 and 1 bar      &  K                   &   Uniform   &   0 &   1000  \\
$\Delta T$ between 1 and 0.1 bar      &  K                   &   Uniform   &   0 &   1000  \\
$\Delta T$ between 0.1 bar and 1 mbar      &  K                   &   Uniform   &   0 &   1000  \\
$\Delta T$ between 1 mbar and 10 nbar      &  K                   &   Uniform   &   0 &   1000  \\
MgSiO$_3$ Mixing Ratio ($\log$(mr$_{\rm{MgSiO}_3}$)) &  \nodata             &   Log-uniform      &  -10       &   -2      \\
Vertical diffusion coefficient ($\log$(K$_{zz}$))           &  cm$2\cdot$s$^{-1}$                 &   Log-uniform       &  5        &   10       \\
$v_{\rm{settling}}/v_{\rm{mixing}}$ ($f_{sed}$)          &  \nodata                 &   Uniform       &  0        &   5       \\
Width of log-normal particle size distribution ($\sigma_g$))           &  \nodata                 &   Uniform       &  1.05        &   3.05       \\
Wavelength shift ($\Delta_\lambda$)         &  $\mu$m              &          Uniform                &   -0.01     &   0.01       \\
y offset ($\Delta_y$)      &  \nodata                   &   Uniform   &   -0.1 &   0.1  \\
Rotational blurring (f$_{\rm{blur}}$)           &  km$\cdot$s$^{-1}$                 &   Uniform       &  2        &   100       \\
\enddata



\end{deluxetable}

\begin{deluxetable}{lccccc}

\tabletypesize{\scriptsize}
\tablewidth{0pt}
\tablecaption{Summary of Retrieval Results.\label{tab:mcmc_result}}
\tablehead{
\colhead{\textbf{Parameter}} &
\colhead{\textbf{Unit}} &
\multicolumn{2}{c}{\textbf{PHOENIX}} &
\multicolumn{2}{c}{\textbf{HR 7672 B}} \\
\colhead{\textbf{Mass?}} &
\colhead{\textbf{}} &
\colhead{\textbf{fixed}} &
\colhead{\textbf{free}} &
\colhead{\textbf{fixed}} &
\colhead{\textbf{free}} 
}


\startdata
$\log$(g)                   &  cgs                
&   $      5.55^{     +0.10}_{     -0.11}$      
&  $      5.24^{     +0.16}_{     -0.20}$   
& $      5.40^{     +0.08}_{     -0.06}$
& $      5.44^{     +0.04}_{     -0.07}$
\\
M$_P$                       &  M$_{\rm{Jupiter}}$  
&   $     72.74^{     +0.66}_{     -0.64}$       
&   $     51.70^{    +23.26}_{    -19.73}$ 
& $     72.84^{     +0.71}_{     -0.72}$ 
& $     89.97^{     +6.86}_{    -11.18}$
\\
$\log$(mr$_{\rm{H}_2\rm{O}}$) &  \nodata             
&   $     -2.72^{     +0.08}_{     -0.08}$     
& $     -2.86^{     +0.12}_{     -0.12}$
& $     -2.69^{     +0.03}_{     -0.03}$
& $     -2.66^{     +0.03}_{     -0.04}$
\\
$\log$(mr$_{\rm{C}\rm{O}}$)       &  \nodata             
&  $     -2.10^{     +0.09}_{     -0.08}$    
&   $     -2.13^{     +0.13}_{     -0.12}$  
& $     -2.46^{     +0.04}_{     -0.04}$
& $     -2.43^{     +0.05}_{     -0.05}$
\\
$\log$(mr$_{\rm{C}\rm{O}_2}$) &  \nodata             
&   $     -6.37^{     +2.30}_{     -2.18}$   
&  $     -6.21^{     +2.25}_{     -2.25}$ 
& $     -6.40^{     +2.37}_{     -2.30}$
& $     -6.45^{     +2.33}_{     -2.29}$
\\
$\log$(mr$_{\rm{C}\rm{H}_4}$) &  \nodata             
&   $     -7.19^{     +1.60}_{     -1.66}$    
&  $     -7.10^{     +1.61}_{     -1.69}$ 
& $     -8.07^{     +1.26}_{     -1.22}$
& $     -7.97^{     +1.31}_{     -1.29}$
\\
t$_{\rm{int}}$      &  K                   
&   $   1627.88^{    +93.33}_{   -102.34}$
&   $   1534.06^{    +95.46}_{    -92.77}$
& $   1530.81^{    +48.28}_{    -28.15}$
& $   1528.32^{    +41.95}_{    -29.13}$
\\
$\Delta T$ between 100 and 32 bar      &  K                   
& $    258.72^{   +247.87}_{   -172.03}$
& $    548.00^{   +658.13}_{   -359.59}$
& $    307.00^{   +314.60}_{   -210.94}$
& $    317.98^{   +326.28}_{   -219.23}$
\\
$\Delta T$ between 32 and 10 bar      &  K                   
& $    103.68^{   +103.15}_{    -69.98}$
& $    210.50^{   +239.61}_{   -141.82}$
& $    235.43^{   +188.47}_{   -140.30}$
& $    204.66^{   +174.18}_{   -127.16}$
\\
$\Delta T$ between 10 and 3.2 bar      &  K                   
& $    111.52^{   +102.02}_{    -72.21}$
& $    141.11^{   +118.29}_{    -89.29}$
& $    333.62^{    +66.59}_{   -100.92}$
& $    311.36^{    +67.24}_{    -95.21}$
\\
$\Delta T$ between 3.2 and 1 bar      &  K                   
& $    391.81^{   +164.72}_{   -156.98}$
& $    214.36^{   +136.42}_{   -118.49}$
& $     27.06^{    +38.12}_{    -19.32}$
& $     33.20^{    +43.91}_{    -23.68}$
\\
$\Delta T$ between 1 and 0.1 bar      &  K                   
& $    498.01^{   +248.72}_{   -243.04}$
& $    482.01^{   +232.47}_{   -219.63}$
& $     18.52^{    +40.72}_{    -13.86}$
& $     20.90^{    +34.02}_{    -15.29}$
\\
$\Delta T$ between 0.1 bar and 1 mbar      &  K                   
& $    485.32^{   +299.40}_{   -289.41}$
& $    496.15^{   +283.12}_{   -287.49}$
& $    871.08^{    +89.54}_{   -143.57}$
& $    828.82^{   +117.63}_{   -179.88}$
\\
$\Delta T$ between 1 mbar and 10 nbar      &  K                   
& $    510.73^{   +286.06}_{   -297.90}$
& $    504.10^{   +287.28}_{   -283.59}$
& $    337.96^{   +360.25}_{   -239.78}$
& $    396.66^{   +359.62}_{   -278.23}$
\\
$\log$(mr$_{\rm{MgSiO}_3}$) &  \nodata             
&  $     -6.47^{     +2.03}_{     -2.12}$
&  $     -4.75^{     +1.73}_{     -2.86}$  
& $     -6.96^{     +1.89}_{     -1.91}$
& $     -6.50^{     +1.97}_{     -2.22}$
\\
$\log$(K$_{zz}$)           &  cm$2\cdot$s$^{-1}$                 
&   $      7.60^{     +1.41}_{     -1.49}$  
&  $      7.59^{     +1.48}_{     -1.48}$
& $      7.65^{     +1.52}_{     -1.67}$
& $      7.62^{     +1.56}_{     -1.69}$
\\
$f_{sed}$          &  \nodata                 
&   $      2.54^{     +1.43}_{     -1.45}$      
&  $      2.86^{     +1.31}_{     -1.51}$
& $      2.52^{     +1.56}_{     -1.57}$
& $      2.56^{     +1.57}_{     -1.60}$
\\
$\sigma_g$           &  \nodata                 
&   $      2.04^{     +0.58}_{     -0.57}$   
&  $      2.03^{     +0.61}_{     -0.57}$   
& $      2.04^{     +0.64}_{     -0.64}$
& $      2.03^{     +0.65}_{     -0.63}$  
\\
$\Delta_\lambda$         &  $\mu$m              
& $      0.00^{     +0.00}_{     -0.00}$  
& $      0.00^{     +0.00}_{     -0.00}$
& $      0.00^{     +0.00}_{     -0.00}$
& $      0.00^{     +0.00}_{     -0.00}$
\\
$\Delta_y$      &  \nodata                   
&   \nodata  
& \nodata 
& $      0.08^{     +0.00}_{     -0.00}$
& $      0.08^{     +0.00}_{     -0.00}$
\\
f$_{\rm{blur}}$           &  km$\cdot$s$^{-1}$                 
&   $   39.62^{   +1.40}_{   -1.39}$       
&  $   38.71^{   +1.67}_{   -1.84}$
& $   45.03^{   +0.45}_{   -0.43}$
& $   45.04^{   +0.55}_{   -0.53}$
\\
$[$C/H$]$ &dex
& $      0.09^{     +0.9}_{     -0.08}$ 
&$      0.06^{     +0.13}_{     -0.11}$ 
& $      -0.27^{     +0.04}_{     -0.04}$
& $      -0.24^{     +0.05}_{     -0.05}$
\\
$[$O/H$]$ &dex
& $      0.00^{     +0.08}_{     -0.07}$
& $      -0.05^{     +0.12}_{     -0.11}$
& $      -0.22^{     +0.04}_{     -0.03}$
& $      -0.19^{     +0.04}_{     -0.04}$
\\
C/O &\nodata
& $      0.72^{     +0.03}_{     -0.03}$
& $      0.77^{     +0.04}_{     -0.04}$
& $      0.52^{     +0.02}_{     -0.02}$
& $      0.52^{     +0.02}_{     -0.02}$
\\
\enddata

\tablecomments{{a: We report the median of posterior distribution and error bars correspond to the difference of the median and the 68\% credible interval. b: We adopt solar elemental abundances from~\citet{Palme2014}.}}

\end{deluxetable}

\end{CJK*}
 
\end{document}
